\pdfoutput=1

\newif\ifbeautified
\beautifiedtrue  

\ifbeautified
  \documentclass[]{fpt}
\else
  \documentclass[11pt]{article}
  \usepackage{acl}
\fi

\usepackage{times}
\usepackage{latexsym}

\usepackage[T1]{fontenc}
\usepackage[utf8]{inputenc}
\usepackage{microtype}

\ifbeautified
  \usepackage{lineno}
  \usepackage[toc,page,header]{appendix}
  \usepackage{minitoc}
  \usepackage{amsfonts}
  \usepackage{tabularx}
  \usepackage{listings}
  \usepackage{xcolor}
  \usepackage{cancel}

  \usepackage{tabulary,multirow,xspace}
  \usepackage{fixmath,mathtools,nicefrac,mmstyle}
  \usepackage{subcaption}
  \usepackage{caption}
  \usepackage{wrapfig}
  \usepackage[misc]{ifsym}
  \usepackage{colortbl}
  \usepackage{multicol}
  \usepackage[most]{tcolorbox}
  \usepackage{pifont}
\fi

\usepackage{paralist}
\usepackage{xspace}
\usepackage{makecell}
\usepackage{booktabs}
\usepackage{amsmath}
\usepackage{amssymb}
\usepackage{pifont}
\usepackage{dblfloatfix}
\usepackage{enumitem}
\usepackage{multirow}
\usepackage{rotating}
\usepackage{array}
\usepackage[autostyle=true]{csquotes}
\usepackage{inconsolata}

\usepackage{colortbl}
\usepackage{xcolor}
\definecolor{Red}{RGB}{192, 0, 0}
\definecolor{Blue}{RGB}{13, 110, 158}
\definecolor{Yellow}{RGB}{218, 169, 20}
\definecolor{lightyellow}{RGB}{255,255,153}

\definecolor{HighlightBlue}{RGB}{0, 100, 148}
\definecolor{HighlightRed}{RGB}{230, 57, 70}

\definecolor{LightRed}{HTML}{ffe0e0}
\definecolor{LightBlue}{HTML}{def5ff}
\definecolor{LightYellow}{HTML}{FFF6DB}
\definecolor{LightGreen}{HTML}{eff9f0}

\definecolor{lightyellow}{RGB}{255,242,204}
\definecolor{lightorange}{RGB}{251,229,214}
\definecolor{lightgreen}{RGB}{226,240,217}
\definecolor{lightblue}{RGB}{222,235,247}
\definecolor{lightgray}{RGB}{209,201,206}
\definecolor{deepgray}{RGB}{178,164,173}
\definecolor{deepblue}{RGB}{70,130,180}

\ifbeautified
  \definecolor{codegreen}{rgb}{0,0.6,0}
  \definecolor{codegray}{rgb}{0.5,0.5,0.5}
  \definecolor{codepurple}{rgb}{0.58,0,0.82}
  \definecolor{backcolour}{rgb}{0.95,0.95,0.92}
  \definecolor{boxblue}{RGB}{13,110,158}
  \definecolor{boxbluebg}{RGB}{215,235,242}
  \definecolor{mygray1}{gray}{.95}
  \definecolor{mygray2}{gray}{.9}
  \definecolor{mygray3}{gray}{.95}
  \definecolor{commentgreen}{rgb}{0.1, 0.4, 0.1}
  \definecolor{keywordblue}{rgb}{0.1, 0.1, 0.7}
  \definecolor{stringred}{rgb}{0.7, 0.1, 0.1}

  \lstdefinestyle{mystyle}{
      backgroundcolor=\color{backcolour},
      commentstyle=\color{codegreen},
      keywordstyle=\color{magenta},
      numberstyle=\tiny\color{codegray},
      stringstyle=\color{codepurple},
      basicstyle=\ttfamily\footnotesize,
      breakatwhitespace=false,
      breaklines=true,
      captionpos=b,
      keepspaces=true,
      numbers=none,
      numbersep=5pt,
      showspaces=false,
      showstringspaces=false,
      showtabs=false,
      tabsize=2
  }
  \lstdefinestyle{pythonstyle}{
      commentstyle=\color{commentgreen},
      keywordstyle=\color{keywordblue},
      stringstyle=\color{stringred},
      basicstyle=\ttfamily\scriptsize,
      breaklines=true,
      keepspaces=true,
      showstringspaces=false,
      frame=none,
      language=Python,
  }
\fi

\usepackage{hyperref}
\usepackage{url}
\usepackage{soul}

\usepackage{wrapfig,lipsum,booktabs}
\usepackage{threeparttable}
\usepackage{graphicx}
\usepackage{adjustbox}
\usepackage{float}
\usepackage{multicol}
\usepackage[ruled,boxed,linesnumbered]{algorithm2e}
\usepackage{caption}
\usepackage{cleveref}
\crefname{figure}{Fig.}{Figs.}
\Crefname{figure}{Fig.}{Figs.}
\crefname{table}{Tab.}{Tabs.}
\Crefname{table}{Tab.}{Tabs.}
\crefname{section}{Sec.}{Secs.}
\Crefname{section}{Sec.}{Secs.}
\crefname{appendix}{Appendix}{Appendix}
\Crefname{appendix}{Appendix}{Appendix}

\newcommand{\cmark}{\textcolor[HTML]{006400}{\ding{51}}}
\newcommand{\xmark}{\textcolor[HTML]{B22222}{\ding{55}}}

\NewDocumentCommand{\revanth}
{ mO{} }{\textcolor{blue}{\textsuperscript{\textit{Revanth}}\textsf{\textbf{\small[#1]}}}}

\ifbeautified

  \newlength\savewidth
  \newcolumntype{x}[1]{>{\centering\arraybackslash}p{#1pt}}

  \newcommand{\app}{\raise.17ex\hbox{$\scriptstyle\sim$}}

  \makeatletter
  \DeclareRobustCommand\onedot{\futurelet\@let@token\@onedot}
  \def\@onedot{\ifx\@let@token.\else.\null\fi\xspace}

  \makeatother

  \makeatletter
  
  \newcommand{\Rmnum}[1]{\expandafter\@slowromancap\romannumeral #1@}
  \makeatother

  \usepackage{graphicx}
\usepackage{amssymb}
\usepackage{pifont}
\usepackage{amsmath}
\usepackage{float}
\usepackage{wrapfig}
\usepackage{multirow}
\usepackage{tcolorbox}
\tcbuselibrary{breakable, skins, raster}
\usepackage{listings}
  \usepackage{setspace}
  \usepackage{tikz}

  \usepackage{booktabs,threeparttable,tabularx,array,ragged2e,xcolor,siunitx}
  \newcolumntype{L}[1]{>{\raggedright\arraybackslash}p{#1}}
\fi

\ifbeautified
  \title{\textcolor{fptred}{CodeWiki}: Evaluating AI's Ability to Generate Holistic Documentation for Large-Scale Codebases}

  \author{
  \centerline{
      Anh Nguyen Hoang $^{1}$\quad
      Minh Le-Anh $^{1}$ \quad
      Bach Le $^{2}$ \quad
      Nghi D.Q. Bui $^{1,{\dagger},{\ddagger}}$
  }%
  \vspace{1mm}%
  \centerline{\small \texttt{\{anhnh2220,minhla4\}@fpt.com \quad bach.le@unimelb.edu.au \quad bdqnghi@gmail.com}}%
  \vspace{-3mm}%
  }

  \affiliation[1]{FPT Software AI Center, Vietnam}
  \affiliation[2]{University of Melbourne, Australia}
  \contribution[\dagger]{Project Lead}
  \contribution[\ddagger]{Corresponding Author}

  \abstract{
  Given a large and evolving codebase, the ability to automatically generate a holistic, architecture-aware documentation that captures not only individual functions but also their cross-file, cross-module, and system-level interactions remains an open challenge. Comprehensive documentation is essential for long-term software maintenance and team collaboration, yet current automated approaches still fail to model the rich semantic dependencies and architectural structures that define real-world software systems. We present \textbf{CodeWiki}, a unified framework for automated repository-level documentation across seven programming languages. CodeWiki introduces three key innovations: (i) hierarchical decomposition that preserves architectural context across multiple levels of granularity, (ii) recursive multi-agent processing with dynamic task delegation for scalable generation, and (iii) multi-modal synthesis that integrates textual descriptions with visual artifacts such as architecture diagrams and data-flow representations. To enable rigorous evaluation, we introduce \textbf{CodeWikiBench}, a comprehensive benchmark featuring multi-dimensional rubrics and LLM-based assessment protocols. Experimental results demonstrate that CodeWiki achieves a 68.79\% quality score with proprietary models, outperforming the closed-source DeepWiki baseline (64.06\%) by 4.73\%, with particularly strong improvements on high-level scripting languages (+10.47\%). We open-source CodeWiki to foster future research and community adoption.
  }

  \checkdata[GitHub]{\url{https://github.com/FSoft-AI4Code/CodeWiki}}
\else
  \title{CodeWiki: Evaluating AI's Ability to Generate Holistic Documentation for Large-Scale Codebases}

  \author{Nguyen Hoang Anh\textsuperscript{1\dag},
  Minh Le-Anh\textsuperscript{1},
  Bach Le\textsuperscript{2},
  Nghi D.Q. Bui\textsuperscript{1,*,\dag} \\
  \\
  \textsuperscript{1}FPT Software AI Center, Vietnam \quad
  \textsuperscript{2}The University of Melbourne, Australia \\
  \texttt{\{anhnh2220,minhla4\}@fpt.com, bach.le@unimelb.edu.au, bdqnghi@gmail.com}
  \\
  \small{*\textit{: Project Lead}},
  \small{\dag\textit{: Corresponding Author}}
  }
\fi

\begin{document}
\maketitle

\ifbeautified
\else
  \begin{abstract}
Given a large and evolving codebase, the ability to automatically generate a holistic, architecture-aware documentation that captures not only individual functions but also their cross-file, cross-module, and system-level interactions remains an open challenge. Comprehensive documentation is essential for long-term software maintenance and team collaboration, yet current automated approaches still fail to model the rich semantic dependencies and architectural structures that define real-world software systems. We present \textbf{CodeWiki}, a unified framework for automated repository-level documentation across seven programming languages. CodeWiki introduces three key innovations: (i) hierarchical decomposition that preserves architectural context across multiple levels of granularity, (ii) recursive multi-agent processing with dynamic task delegation for scalable generation, and (iii) multi-modal synthesis that integrates textual descriptions with visual artifacts such as architecture diagrams and data-flow representations. To enable rigorous evaluation, we introduce \textbf{CodeWikiBench}, a comprehensive benchmark featuring multi-dimensional rubrics and LLM-based assessment protocols. Experimental results demonstrate that CodeWiki achieves a 68.79\% quality score with proprietary models, outperforming the closed-source DeepWiki baseline (64.06\%) by 4.73\%, with particularly strong improvements on high-level scripting languages (+10.47\%). We open-source CodeWiki to foster future research and community adoption.

\end{abstract}

\fi

\section{Introduction}

\begin{figure*}[t]
    \centering
    \includegraphics[width=\textwidth]{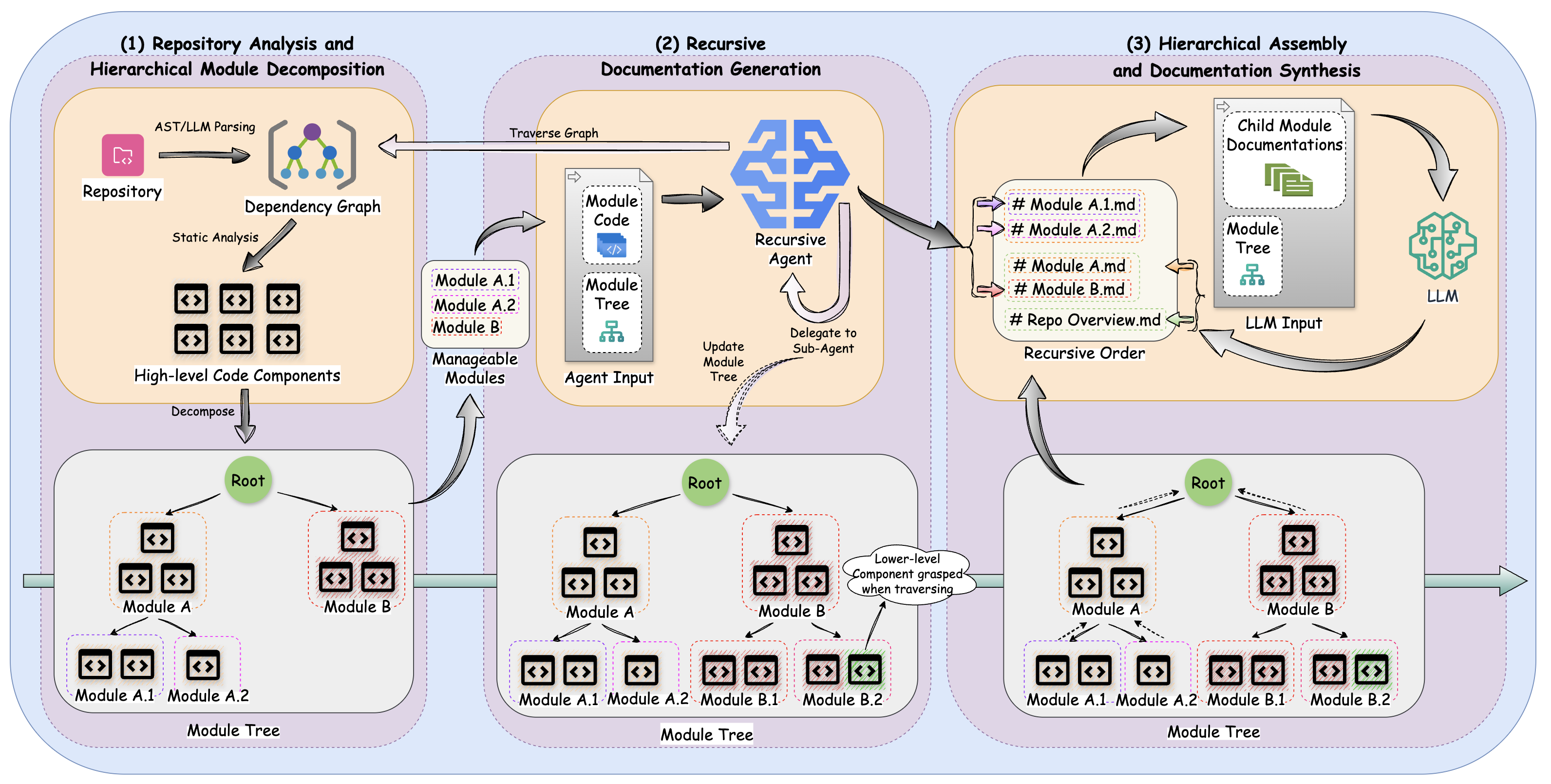}
    \caption{CodeWiki Framework Architecture Overview. The framework operates in three main phases: (1) Repository analysis through AST/LLM parsing to construct dependency graphs and identify high-level components, followed by hierarchical decomposition into manageable modules; (2) Recursive documentation generation where specialized agents process leaf modules with dynamic delegation capabilities, creating markdown documentation while maintaining cross-module references; (3) Hierarchical assembly where parent modules are synthesized from child documentation using LLM-based synthesis, culminating in comprehensive repository overview documentation. The module tree evolves throughout the process, enabling scalable processing of repositories of arbitrary size.}
    \label{fig:framework-architecture}
\end{figure*}

In the rapidly evolving landscape of software development, maintaining comprehensive and up-to-date documentation has become increasingly challenging as codebases grow in size and complexity. Recent industry surveys reveal that documentation represents a critical bottleneck in modern development workflows: approximately 31\% of developers currently rely heavily on AI for documenting code, while 25\% use AI extensively for creating and maintaining documentation—making documentation tasks among the highest priorities for AI integration over the next 3-5 years~\citep{stackoverflow2025}. Notably, while developers express high satisfaction with AI tools for searching and learning, they indicate plans to substantially increase AI adoption for documentation and testing tasks, suggesting current tools remain insufficient for these critical needs. This widespread focus on AI-driven documentation reflects its dual burden: while essential for software maintenance and collaboration~\citep{souza2005astudyofthedocumentation,zhi2015costbenefitquality,chenandhuang2009}, manual creation and maintenance remain labor-intensive and often neglected~\citep{mcburneyetal-documentationeffort-2018}. The emergence of Large Language Models (LLMs)~\citep{Zhang2023ASO,he2025llmbasedmultiagent} has opened opportunities for automating documentation generation. However, current approaches still face two fundamental limitations.

\paragraph{\textbf{Limitations of Scalability at the Repository Level}}
Recent advances in code understanding and natural language generation have demonstrated promising results in function-level and file-level documentation tasks~\citep{feng-etal-2020-codebert,khan2023automaticcodedocumentation, poudel2024documint, makharev2025beyondfunctionlevel, lomshakov-etal-2024-proconsul,fewshot-code-summarization-ase-2022, readsum-2023,luo-etal-2024-repoagent, yang-etal-2025-docagent}
However, existing approaches face significant challenges when scaling to repository-level documentation.

Unlike function-level documentation that focuses on individual components, repository-level documentation task is to capture architectural patterns, cross-module interactions, data flows, and system-wide design decisions~\citep{rai2022areviewonsourcecodedocumentation,treude2020beyondaccuracy}. Furthermore, the hierarchical nature of software systems requires documentation that can serve multiple audiences, from high-level stakeholders seeking architectural overviews to developers requiring detailed implementation guidance~\citep{nassif2025non-linear}.

$\hookrightarrow$ To address these challenges, we present \textbf{CodeWiki}, an open-source framework specifically designed to generate holistic, structured repository-level documentation that includes diverse content types such as system architecture diagrams, data flow visualizations, and sequence diagrams while scaling effectively for projects of arbitrary size. CodeWiki introduces several key innovations: a hierarchical decomposition strategy inspired by dynamic programming principles that breaks complex repositories into manageable modules while preserving architectural coherence, a recursive agentic system with dynamic delegation capabilities that enables adaptive processing based on module complexity, and a comprehensive documentation synthesis approach that generates multiple forms of visual and textual documentation reflecting fine-grained codebase collaboration patterns.

\paragraph{\textbf{Limitations of the Evaluation Landscape}}
Despite progress, systematic benchmarks remain lacking. Function-level and file-level work relies on general-purpose metrics (BLEU, ROUGE) that fail to capture documentation quality nuances~\citep{royetal-reassessing-2021, evtikhiev2023outofbleu}. Repository-level evaluation is particularly challenging as documentation admits multiple valid structures sharing underlying architectural content, highlighting the need for robust assessment beyond surface-level similarity.

$\hookrightarrow$ To fill this gap, we introduce \textbf{CodeWikiBench}, a comprehensive benchmark specifically designed for repository-level documentation. Instead of relying on surface-level similarity, we derive hierarchical rubrics from the official documentation maintained by project developers, enabling the evaluation of underlying repository properties. For robustness, CodeWikiBench further adopts an agentic assessment framework in which judge agents evaluate generated documentation against rubric requirements and aggregate results hierarchically, producing both quality scores and reliability measures.

\paragraph{\textbf{Limitations of Multilingual Support}}
Most studies on code documentation generation have primarily focused on Python, with limited consideration of other widely used programming languages such as Java, JavaScript, C, or C++. This narrow scope restricts generalizability and overlooks the structural diversity inherent in real-world software projects, where repositories often involve multiple languages.

$\hookrightarrow$ CodeWiki supports repository-level documentation generation across 7 programming languages (Python, Java, JavaScript, TypeScript, C, C++ and C\#). In addition, CodeWikiBench provides a multilingual benchmark built from multiple repositories in the same set of languages, enabling comprehensive evaluation.

Using CodeWikiBench, we evaluate the effectiveness of CodeWiki against multiple repository-level documentation systems, including open-source baselines (\textbf{OpenDeepWiki}\footnote{\url{https://github.com/AIDotNet/OpenDeepWiki}}, \textbf{deepwiki-open}\footnote{\url{https://github.com/AsyncFuncAI/deepwiki-open}}) and the closed-source \textbf{DeepWiki}. The results show that CodeWiki achieves an overall quality score of 68.79\%, outperforming all baselines with improvements of up to 18.54\% over the strongest baseline, DeepWiki, on individual repositories. CodeWiki produces comprehensive documentation encompassing architectural diagrams, usage patterns, and cross-module dependency visualizations (see Appendix~\ref{app:documentation-example} for an example).

In summary, the main contributions of this work are:
\begin{enumerate}[leftmargin=*]
    \item \textbf{CodeWiki}, an open-source semi-agentic framework for scalable repository-level documentation with proprietary and open-source LLM backends.
    \item \textbf{CodeWikiBench}, a comprehensive repository-level documentation benchmark with hierarchical rubrics and agentic assessment.
    \item Comprehensive multilingual repository documentation supporting 7 widely used languages.
    \item Experimental validation demonstrating improvements over all baselines, with gains of up to 18.54\% on individual repositories and 4.73\% on average compared to the closed-source DeepWiki system.
\end{enumerate}

\noindent A detailed discussion of related work and positioning relative to prior approaches is provided in Appendix~\ref{app:related-work}.

\ifbeautified
  \section{Related Work}

\subsection{Code Intelligence and LLMs}

The rapid advancement of Large Language Models (LLMs) has revolutionized code intelligence tasks. Early work by von der Mosel et al. \cite{Mosel} demonstrated the effectiveness of Transformer-based models for software engineering, leading to specialized code models like CodeBERT \cite{feng-etal-2020-codebert}, GraphCodeBERT \cite{guo2021graphcodebert}, CodeT5 \cite{wang2021codet5}, CodeT5+ \cite{wang2023codet5plus}, and CodeTF \cite{bui2023codetf}. These established early transformer-based approaches for code understanding and generation.

Comprehensive surveys by Niu et al. \cite{niu2022deep} and Hou et al. \cite{hou2023large} established the importance of domain-specific pre-training, showing consistent improvements over general-purpose language models across code search, summarization, and generation tasks. However, recent studies highlight important limitations: Li et al. \cite{Liyao} found that models often make confident decisions without sufficient reasoning, while Zheng et al. \cite{zheng2023understanding} noted that LLM performance remains unstable and requires careful prompt engineering \cite{shin2023prompt}.

For code generation, HumanEval \cite{chen2021evaluating} and MBPP \cite{mbpp} established foundational benchmarks, later extended to multiple languages by MultiPL-E \cite{mul-e} and CodeScope \cite{yan2023codescope}. More recently, CodeMMLU \cite{nguyencodemmlu} has emerged as a comprehensive multi-task benchmark for assessing code understanding and reasoning capabilities of CodeLLMs.

\subsection{Automated Documentation Generation}

Documentation generation has emerged as a critical LLM application in software engineering. Traditional approaches focused on function-level summarization using datasets like CodeSearchNet \cite{husain2019codesearchnet} and FunCom \cite{leclair2019neural}. While Sun et al. \cite{sun2023automatic} found that ChatGPT can generate coherent comments, their quality often falls short of specialized models when evaluated using BLEU and ROUGE metrics.

Recent research has shifted toward repository-level documentation generation, which requires understanding complex dependencies and architectural context. Luo et al. \cite{luo-etal-2024-repoagent} introduced RepoAgent, an LLM-powered framework that leverages static analysis and incremental processing. DocAgent \cite{yang-etal-2025-docagent} proposed a multi-agent system with specialized roles (Reader, Searcher, Writer, Verifier) for collaborative documentation generation. Both systems demonstrate significant improvements over baseline approaches but struggle with scaling to large repositories and maintaining cross-file consistency.

Evaluation methodologies have evolved from traditional NLP metrics like BLEU \cite{papineni2002bleu} and ROUGE \cite{lin2004rouge} to more sophisticated approaches. Roy et al. \cite{roy2021reassessing} and Mastropaolo et al. \cite{mastropaolo2024evaluating} questioned the effectiveness of automatic metrics, while recent work has adopted LLM-as-a-judge methodologies \cite{zheng2023judging, liu2023geval}. Multi-dimensional evaluation frameworks now assess completeness, helpfulness, and truthfulness, recognizing that documentation quality cannot be reduced to a single dimension.

\subsection{Multi-Agent Systems for Software Engineering}

Multi-agent architectures have gained prominence for tackling complex software engineering tasks requiring diverse expertise and collaborative problem-solving. AutoGen \cite{wu2023autogen} introduced a framework enabling multi-agent communication with language models, allowing agents to assume different roles and collaborate on complex tasks. MetaGPT \cite{hong2023metagpt} demonstrated meta-programming for collaborative frameworks, where agents follow software development methodologies to produce comprehensive solutions.

Multi-agent approaches include HyperAgent \cite{phan2024hyperagent}, which presents generalist software engineering agents capable of solving coding tasks at scale, and AgileCoder \cite{nguyen2025agilecoder}, which introduces dynamic collaborative agents for software development based on agile methodology. In the code generation domain, MapCoder \cite{zhang2023mapcoder} employed map-reduce style collaboration to break down complex programming problems, while ChatDev \cite{qian2023chatdev} showed that multi-agent systems can manage the entire software development lifecycle. The integration of multi-agent systems with documentation represents a recent innovation, with DocAgent demonstrating how role-based collaboration can improve documentation quality through specialized agents for analysis, information retrieval, drafting, and verification. However, existing approaches often lack systematic dependency management and struggle with maintaining consistency across large codebases.
\fi
\section{Our Approach - CodeWiki}

CodeWiki is a semi-agentic framework that automatically generates comprehensive repository-level documentation by addressing context limitations through hierarchical decomposition. As illustrated in Figure~\ref{fig:framework-architecture}, our approach operates through three phases: repository analysis and module decomposition, recursive documentation generation, and hierarchical assembly, inspired by dynamic programming principles.

\subsection{Repository Analysis and Hierarchical Module Decomposition}

We begin with static analysis to construct a dependency graph capturing structural relationships, crucial for understanding architecture and identifying decomposition boundaries.

\paragraph{Dependency Graph Construction}
Following DocAgent~\citep{yang-etal-2025-docagent}, we employ Tree-Sitter parsers to extract ASTs, systematically identifying functions, methods, classes, structs, modules, and interdependencies: function calls, class inheritance, attribute access, and module imports. We normalize these to a unified \texttt{depends\_on} relation for cross-language generalization, creating directed graph $G = (V, E)$ where edge $(u, v) \in E$ indicates component $u$ depends on $v$~\citep{liu2023learninggraphbased}.

\paragraph{Entry Point Identification and Decomposition}
Topological sorting identifies zero-in-degree components—independent entry points where users interact (main functions, API endpoints, CLI, public interfaces). High-level components are hierarchically decomposed into manageable modules through recursive partitioning considering component interdependencies and semantic coherence. For scalability, only component IDs serve as input, resulting in a feature-oriented module tree.

\subsection{Recursive Documentation Generation}

CodeWiki's core innovation is recursive agent-based processing, enabling arbitrary repository size handling while maintaining bounded complexity and architectural coherence.

\paragraph{Agent Architecture}
Each leaf module is assigned a specialized agent equipped with: (1) complete source code access, (2) full module tree for cross-module understanding, (3) documentation workspace tools (view, create, edit operations), and (4) dependency graph traversal for contextual exploration.

The agent workflow: analyzes components to understand functionality and interfaces, explores context by traversing dependencies within and across modules, and generates comprehensive markdown documentation including descriptions, usage examples, API specifications, and architectural insights.

\paragraph{Dynamic Delegation}
Our key innovation is adaptive scalability through dynamic delegation. When module complexity exceeds single-pass capacity, agents delegate sub-modules to specialized sub-agents. Delegation criteria: code complexity metrics (cyclomatic complexity, nesting depth), semantic diversity (functionally distinct subcomponents), and context window utilization.

The recursive process follows bottom-up processing (see Algorithm~\ref{alg:recursive-doc} in Appendix~\ref{app:algorithm}). Upon delegation, agents provide sub-module specifications, the module tree updates, and newly created leaves process recursively. This handles modules of any size while maintaining quality.

\paragraph{Cross-Module Reference Management}
Maintaining coherence across boundaries requires sophisticated reference management. When agents encounter external components during traversal, our intelligent resolution system creates cross-references rather than duplicating content. A global registry tracks documented components and locations, enabling quick identification of documented components.

This ensures conciseness while preserving completeness, creating interconnected documentation reflecting the actual codebase structure, enabling easy navigation between related components.

\subsection{Hierarchical Assembly and Documentation Synthesis}

After leaf documentation, hierarchical assembly synthesizes component-level details into architectural overviews through recursive parent module processing.

Parent modules undergo LLM synthesis with carefully crafted prompting. LLMs receive: child module documentation, module tree structure, dependency information, and synthesis instructions for architectural patterns and feature interactions.

Synthesis involves multiple stages: analyzing child documentation for themes and patterns, generating architectural overviews explaining module collaboration, creating feature summaries distilling capabilities, developing usage guides for public interfaces, and producing architectural diagrams visualizing relationships and data flows.

\section{Benchmarking System - CodeWikiBench}

\begin{figure}[t]
    \centering
    \includegraphics[width=0.85\columnwidth]{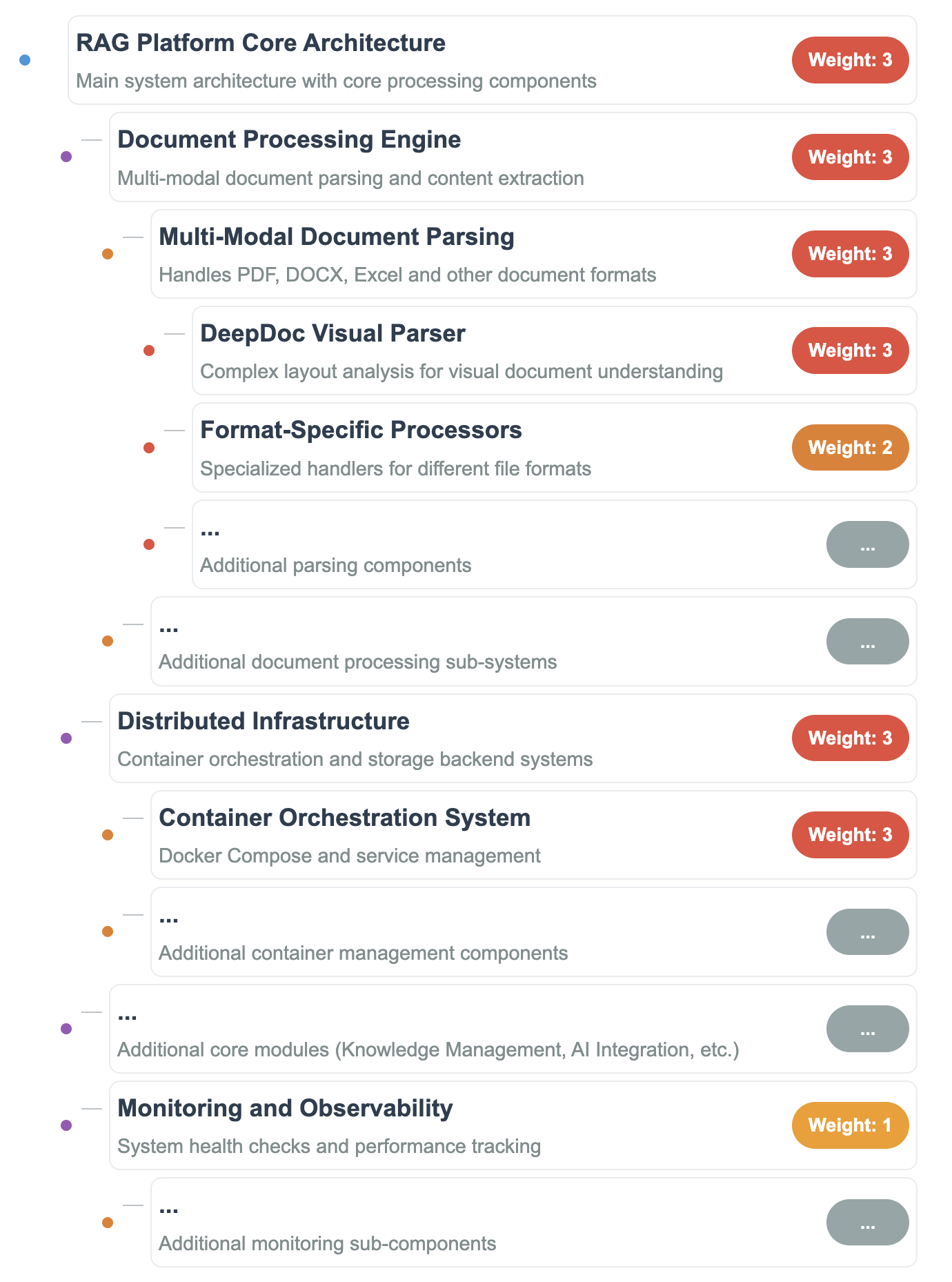}
    \caption{Example hierarchical evaluation rubric for the RAGFlow repository. The rubric mirrors the project's architectural structure with weighted requirements at multiple levels. Leaf nodes represent specific requirements assessed by Judge Agent, while parent scores are computed through weighted aggregation. The hierarchical organization ensures comprehensive coverage from high-level architectural components down to specific implementation details, with weights reflecting the relative importance of each component for understanding the repository}
    \label{fig:rubric-ragflow}
\end{figure}

Repository-level documentation evaluation presents unique challenges due to complexity and hierarchical organization. Unlike code generation assessable through automated metrics, documentation quality requires nuanced evaluation of completeness, accuracy, and coherence across abstraction levels. Recent research highlights traditional metrics' limitations for code tasks~\citep{evtikhiev2023outofbleu,haldar-hockenmaier-2024-analyzing}, while LLM-as-a-Judge paradigms offer new possibilities~\citep{wang2025canllmsreplacehumanevaluator,crupi2025effectivenessllmasajudgecodegeneration,he2025codecourtroomllmsnew}. We develop a comprehensive framework synthesizing repository-specific rubrics and employing agentic assessment.

\subsection{Hierarchical Rubric Generation}

Following~\citep{starace2025paperbenchevaluatingaisability}, we synthesize evaluation rubrics tailored for each repository, illustrated in Figure~\ref{fig:rubric-ragflow}. We collect official documentation from selected open-source projects and hierarchically parse into structured JSON format, leveraging directory structure and markdown syntax. This captures multi-level technical documentation from architectural overviews to implementation details.

A Rubric Generator Agent processes each structure with tool access to detailed contents, generating hierarchical rubrics mirroring repository functionality and aligning with actual documentation standards. To enhance reliability, we employ multiple independent generations using different model families. Final rubrics emerge from synthesizing these perspectives, reducing single-model biases. Analysis shows generated rubrics demonstrate high consistency: 73.65\% semantic reliability and 70.84\% structural reliability (Appendix~\ref{app:rubric-reliability}).

\subsection{Documentation Assessment}

Our evaluation employs specialized Judge Agents operating under carefully designed protocols. Judges receive generated documentation in structured JSON format with hidden detailed contents, plus repository-specific rubrics. The systematic methodology: analyzes documentation structure and content, searches comprehensively for requirement coverage, makes binary adequacy decisions, and provides concise reasoning.

Critically, judges evaluate only leaf-level requirements, ensuring assessment grounds in specific concrete criteria rather than abstract concepts. For example, in Figure~\ref{fig:rubric-ragflow}, agents assess "DeepDoc Visual Parser for complex layouts" rather than broader "Document Processing Engine."

For enhanced reliability, multiple Judge Agents using different model families evaluate each requirement. Final leaf scores average across assessments, reducing individual model biases.

\subsection{Hierarchical Score Aggregation}

Final quality scores compute through bottom-up weighted aggregation respecting rubric hierarchy while tracking reliability via standard deviation propagation. Let $R$ be the rubric root, $C(n)$ denote node $n$'s children, and $w(n)$ denote assigned weight.

\paragraph{Leaf Assessment and Reliability}
For leaf nodes $\ell$, multiple assessments $s_1, s_2, \ldots, s_m$ aggregate as:
$$S(\ell) = \bar{s} = \frac{1}{m}\sum_{i=1}^{m}s_i$$

Reliability quantified via standard deviation:
$$\sigma_{\ell} = \sqrt{\frac{1}{m-1}\sum_{i=1}^{m}(s_i - \bar{s})^2}$$

Lower deviations indicate higher judge consensus and assessment reliability.

\paragraph{Hierarchical Propagation}
For internal nodes $n$ with children $C(n) = \{c_1, \ldots, c_k\}$ having scores $S(c_i)$, weights $w(c_i)$, and deviations $\sigma_{c_i}$:
$$S(n) = \frac{\sum_{i=1}^{k} w(c_i) \cdot S(c_i)}{\sum_{i=1}^{k} w(c_i)}, \quad
\sigma_n = \frac{\sqrt{\sum_{i=1}^{k} w(c_i)^2 \sigma_{c_i}^2}}{\sum_{i=1}^{k} w(c_i)}$$
This ensures uncertainty propagates appropriately, with weights determining score contribution and reliability influence.

\paragraph{Final Assessment}
Final repository score $S(R) \in [0, 1]$ with deviation $\sigma_R$ represents comprehensive weighted aggregation propagating assessments and reliability upward. Lower $\sigma_R$ indicates higher methodology reliability. This dual-metric approach provides quality assessment with confidence bounds, enabling nuanced interpretation. Weights reflect component importance based on criticality and complexity while maintaining statistical rigor in uncertainty quantification.

\begin{table*}[!ht]
    \centering
    \resizebox{.95\textwidth}{!}{%
    \begin{tabular}{llrrccc}
        \hline
        \textbf{Repository} & \textbf{Language} & \textbf{LOC} & \textbf{System} & \textbf{Score (\%)} & \textbf{Coverage} & \textbf{Improvement (\%)} \\
        \hline
        \multirow{4}{*}{All-Hands-AI--OpenHands} & \multirow{4}{*}{Python} & \multirow{4}{*}{229,909} & OpenDeepWiki & 58.12 ± 3.21 & 42/67 & -- \\
         & & & deepwiki-open & 61.35 ± 2.98 & 45/67 & -- \\
         & & & DeepWiki & 73.04 ± 2.54 & 54/67 & -- \\
         & & & \cellcolor[HTML]{CBCEFB}CodeWiki & \cellcolor[HTML]{CBCEFB}\textbf{82.45 ± 2.65} & \cellcolor[HTML]{CBCEFB}\textbf{59/67} & \cellcolor[HTML]{CBCEFB}\textbf{+9.41} \\
        \hline
        \multirow{4}{*}{sveltejs--svelte} & \multirow{4}{*}{JavaScript} & \multirow{4}{*}{124,576} & OpenDeepWiki & 55.23 ± 3.85 & 61/96 & -- \\
         & & & deepwiki-open & 57.89 ± 3.62 & 64/96 & -- \\
         & & & DeepWiki & 68.51 ± 3.31 & 76/96 & -- \\
         & & & \cellcolor[HTML]{CBCEFB}CodeWiki & \cellcolor[HTML]{CBCEFB}\textbf{71.96 ± 3.73} & \cellcolor[HTML]{CBCEFB}\textbf{80/96} & \cellcolor[HTML]{CBCEFB}\textbf{+3.45} \\
        \hline
        \multirow{4}{*}{puppeteer--puppeteer} & \multirow{4}{*}{TypeScript} & \multirow{4}{*}{136,302} & OpenDeepWiki & 51.82 ± 4.15 & 48/82 & -- \\
         & & & deepwiki-open & 54.67 ± 3.94 & 51/82 & -- \\
         & & & DeepWiki & 64.46 ± 3.72 & 60/82 & -- \\
         & & & \cellcolor[HTML]{CBCEFB}CodeWiki & \cellcolor[HTML]{CBCEFB}\textbf{83.00 ± 3.37} & \cellcolor[HTML]{CBCEFB}\textbf{74/82} & \cellcolor[HTML]{CBCEFB}\textbf{+18.54} \\
        \hline
        \multirow{4}{*}{Unity-Technologies--ml-agents} & \multirow{4}{*}{C\#} & \multirow{4}{*}{86,106} & OpenDeepWiki & 62.45 ± 4.28 & 32/46 & -- \\
         & & & deepwiki-open & 65.12 ± 4.05 & 34/46 & -- \\
         & & & DeepWiki & 74.80 ± 3.69 & 39/46 & -- \\
         & & & \cellcolor[HTML]{CBCEFB}CodeWiki & \cellcolor[HTML]{CBCEFB}\textbf{79.78 ± 5.02} & \cellcolor[HTML]{CBCEFB}\textbf{42/46} & \cellcolor[HTML]{CBCEFB}\textbf{+4.98} \\
        \hline
        \multirow{4}{*}{elastic--logstash} & \multirow{4}{*}{Java} & \multirow{4}{*}{117,485} & OpenDeepWiki & 41.25 ± 4.52 & 28/57 & -- \\
         & & & deepwiki-open & 44.18 ± 4.31 & 31/57 & -- \\
         & & & DeepWiki & 54.80 ± 4.10 & 38/57 & -- \\
         & & & \cellcolor[HTML]{CBCEFB}CodeWiki & \cellcolor[HTML]{CBCEFB}\textbf{57.90 ± 3.43} & \cellcolor[HTML]{CBCEFB}\textbf{38/57} & \cellcolor[HTML]{CBCEFB}\textbf{+3.10} \\
        \hline
        \multirow{4}{*}{wazuh--wazuh} & \multirow{4}{*}{C} & \multirow{4}{*}{1,446,730} & OpenDeepWiki & 32.56 ± 5.82 & 18/46 & -- \\
         & & & deepwiki-open & 35.89 ± 5.45 & 21/46 & -- \\
         & & & DeepWiki & \textbf{68.68 ± 4.74} & \textbf{39/46} & -- \\
         & & & \cellcolor[HTML]{CBCEFB}CodeWiki & \cellcolor[HTML]{CBCEFB}64.17 ± 5.44 & \cellcolor[HTML]{CBCEFB}34/46 & \cellcolor[HTML]{CBCEFB}-4.51 \\
        \hline
        \multirow{4}{*}{electron--electron} & \multirow{4}{*}{C++} & \multirow{4}{*}{184,234} & OpenDeepWiki & 28.45 ± 3.95 & 35/92 & -- \\
         & & & deepwiki-open & 31.22 ± 3.78 & 38/92 & -- \\
         & & & DeepWiki & \textbf{44.10 ± 3.12} & \textbf{54/92} & -- \\
         & & & \cellcolor[HTML]{CBCEFB}CodeWiki & \cellcolor[HTML]{CBCEFB}42.30 ± 3.26 & \cellcolor[HTML]{CBCEFB}48/92 & \cellcolor[HTML]{CBCEFB}-1.80 \\
        \hline
        \hline
        \multirow{4}{*}{\textbf{Average}} & & & OpenDeepWiki & 47.13 ± 4.25 & & -- \\
         & & & deepwiki-open & 50.05 ± 4.02 & & -- \\
         & & & DeepWiki & 64.06 ± 3.60 & & -- \\
         & & & \cellcolor[HTML]{CBCEFB}CodeWiki & \cellcolor[HTML]{CBCEFB}\textbf{68.79 ± 3.84} & & \cellcolor[HTML]{CBCEFB}\textbf{+4.73} \\
        \hline
    \end{tabular}}
    \vspace{+2mm}
    \caption{Documentation generation performance comparison across repositories. Scores represent weighted averages with standard deviations indicating assessment consensus. Coverage shows satisfied leaf criteria out of total requirements. OpenDeepWiki, deepwiki-open are open-source alternatives; DeepWiki is the closed-source baseline.}
    \label{tab:results}
    \vspace{-1mm}
\end{table*}
\section{Experiment}
We conduct comprehensive experiments to evaluate CodeWiki's effectiveness in generating repository-level documentation across diverse programming languages and project types. Our evaluation is structured around three core research questions that address the fundamental challenges of automated repository documentation generation.

\subsection{Research Questions}

Our experimental evaluation is guided by the following research questions:

\textbf{RQ1: Documentation Quality and Coverage} 
How does CodeWiki compare to existing state-of-the-art repository documentation systems in terms of overall documentation quality and requirement coverage?

\textbf{RQ2: Cross-Language Generalization} 
Does CodeWiki demonstrate consistent performance across diverse programming languages, particularly comparing high-level scripting languages (Python, JavaScript, TypeScript) versus systems programming languages (C, C++, C\#, Java)?

\textbf{RQ3: Scalability and Reliability} 
How does CodeWiki's hierarchical decomposition approach perform across repositories of varying sizes and complexity, and what is the reliability of our evaluation methodology?

\subsection{Baselines}

We compare CodeWiki against multiple repository-level documentation systems spanning both open-source and closed-source solutions:

\paragraph{Open-Source Baselines}
\textbf{OpenDeepWiki} and \textbf{deepwiki-open} are open-source alternatives applying LLMs to entire repositories for documentation generation. These systems represent the current state of open-source repository-level documentation tools.

\paragraph{Closed-Source Baseline}
\textbf{DeepWiki}~\citep{deepwiki} is a commercial system has demonstrated effectiveness in industrial applications for automated repository-level documentation generation.

\paragraph{Positioning Relative to Function-Level Approaches}
We note that function-level documentation systems such as RepoAgent~\citep{luo-etal-2024-repoagent} and DocAgent~\citep{yang-etal-2025-docagent} generate N separate documents for N components, representing a fundamentally different documentation paradigm from CodeWiki's hierarchically synthesized module-level outputs. Direct quantitative comparison would require forcing one approach into evaluation criteria it was not designed to satisfy, so we focus our comparison on systems targeting similar repository-level documentation goals.

\subsection{Experimental Setup}

\paragraph{Repository Selection}

We curate a benchmark of 7 open-source repositories spanning diverse programming languages, project scales, and application domains. Our selection prioritizes: (1) language diversity for cross-language generalization, (2) varying codebase sizes ranging from 86K to 1.4M LOC, (3) actively maintained projects with high-quality official documentation for rubric construction, and (4) coverage of multiple domains such as web frameworks, automation tools, and machine learning platforms.

Table~\ref{tab:results} presents the repository characteristics, including primary languages and codebase sizes. The benchmark covers JavaScript (svelte), TypeScript (puppeteer), C (wazuh), C++ (electron), C\# (ml-agents), Java (logstash), and Python (OpenHands), ensuring comprehensive coverage across major programming paradigms.

\paragraph{Temporal Separation for Data Leakage Prevention}

To ensure evaluation integrity, we use repository snapshots from August-September 2025, which postdate the knowledge cutoffs of all models used in our experiments (Claude Sonnet 4: March 2025). This temporal gap provides strong evidence against data leakage concerns. Detailed commit information is provided in Appendix~\ref{app:temporal-separation}.

\paragraph{Model Configuration}

Our experimental configuration employs state-of-the-art language models: Claude Sonnet 4 for documentation generation, multiple models (Claude Sonnet 4, Gemini 2.5 Pro, and Kimi K2 Instruct) for rubric generation to reduce single-model bias, and three distinct judge models (Gemini 2.5 Flash, GPT OSS 120B, and Kimi K2 Instruct) for independent assessments ensuring robust evaluation through multi-model consensus. Configuration details are in Appendix~\ref{app:implementation}.

\paragraph{Evaluation Protocol}

For each repository, we execute the CodeWiki pipeline and generate comprehensive documentation including architectural overviews, component descriptions, and usage guides. The evaluation process follows our hierarchical rubric methodology: (1) generating repository-specific rubrics from official documentation, (2) performing agentic assessment of generated documentation, and (3) computing weighted scores with uncertainty quantification through standard deviation tracking.

\subsection{Results and Analysis}

\subsubsection{RQ1: Documentation Quality and Coverage}

Table~\ref{tab:results} presents comprehensive evaluation results addressing the fundamental question of documentation quality. CodeWiki achieves an average documentation quality score of 68.79\%, representing a 4.73 percentage point improvement over the closed-source DeepWiki baseline (64.06\%) and substantial improvements over open-source alternatives OpenDeepWiki (47.13\%) and deepwiki-open (50.05\%). This demonstrates the effectiveness of our hierarchical synthesis approach compared to both simpler open-source methods and commercial solutions.

The coverage metrics reveal that CodeWiki consistently satisfies more leaf-level requirements than baselines in most cases. Notably, CodeWiki outperforms all baselines in 5 out of 7 repositories, with particularly strong improvements in TypeScript (83.00\% vs. DeepWiki's 64.46\%), Python (82.45\% vs. 73.04\%), and JavaScript (71.96\% vs. 68.51\%). The open-source alternatives show notably degraded performance on more complex codebases, suggesting that simpler whole-repository prompting approaches do not scale effectively.

\textbf{Answer:} CodeWiki demonstrates superior overall documentation quality and coverage compared to both open-source and closed-source systems, achieving improvements of 21.66\% over OpenDeepWiki, 18.74\% over deepwiki-open, and 4.73\% over DeepWiki.

\subsubsection{RQ2: Cross-Language Generalization}

\begin{figure*}[!htbp]
    \centering
    \includegraphics[width=\textwidth]{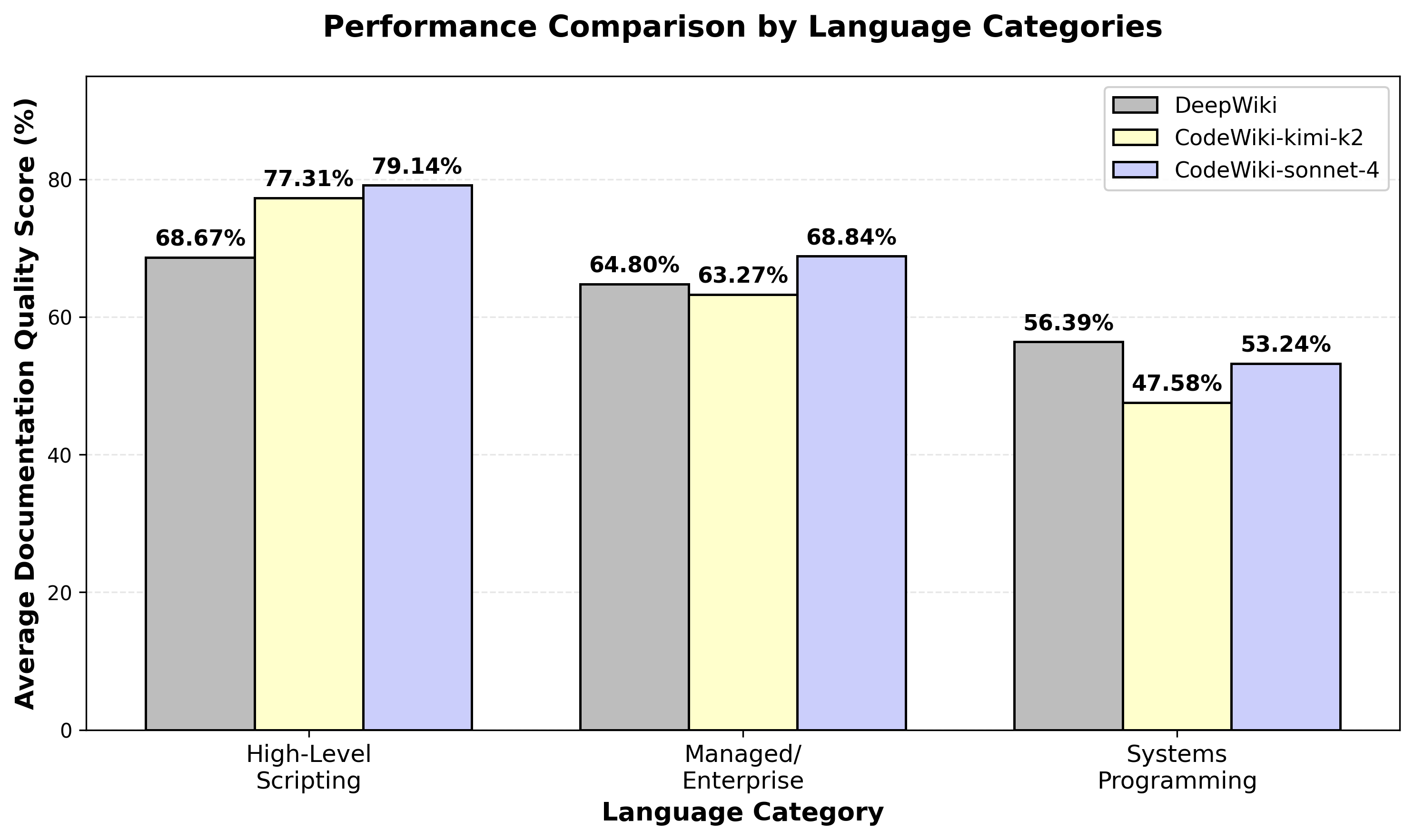}
    \caption{Performance comparison by programming language categories. CodeWiki shows strong improvements in high-level scripting languages (+10.47\%) and managed/enterprise languages (+4.04\%), while systems programming languages present challenges for both systems.}
    \label{fig:language_comparison}
\end{figure*}

Our analysis reveals distinct performance patterns across programming language categories, as visualized in Figure~\ref{fig:language_comparison}.

CodeWiki demonstrates strong performance on scripting languages, achieving average scores of 79.14\% (Python, JavaScript, TypeScript) repositories - a 10.47\% improvement over DeepWiki's 68.67\% average for the same languages.

For managed languages (C\#, Java), CodeWiki achieves competitive performance with moderate but consistent improvements of 4.98\% and 3.10\% respectively, averaging 68.84\% versus DeepWiki's 64.80\%. Systems programming languages (C, C++) present challenges where both frameworks exhibit similar patterns: DeepWiki achieves 56.39\% while CodeWiki reaches 53.24\%. This suggests that specialized parsing strategies may benefit both approaches for handling complex constructs like pointer and template metaprogramming.

\textbf{Answer:} CodeWiki demonstrates strong generalization with notable performance on scripting languages (79.14\% average) and consistent improvements on managed languages, while systems programming languages present similar challenges for both frameworks.

\subsubsection{RQ3: Scalability and Reliability}

Repository size analysis demonstrates CodeWiki's scalability across diverse project scales. Our framework processes repositories ranging from 86K LOC to 1.4M LOC, with performance consistency observed within language categories regardless of repository size. For high-level languages, CodeWiki maintains strong performance across different scales: Unity ML-Agents (86K LOC) achieves 79.78\%, while the larger OpenHands (230K LOC) reaches 82.45\%. This consistency demonstrates that our hierarchical approach successfully addresses context limitation challenges that typically constrain documentation generation at scale.

Critically, both CodeWiki and DeepWiki exhibit similar behavioral patterns across different repository sizes within the same programming language families. Systems programming languages present consistent challenges for both frameworks regardless of project scale, suggesting that the primary factor affecting documentation quality is language-specific parsing complexity rather than repository size. The dynamic delegation mechanism in CodeWiki automatically adapts to varying complexity levels, maintaining bounded processing requirements through hierarchical decomposition.

Reliability analysis through standard deviation measurements validates our multi-model assessment methodology. The average standard deviations of 3.84\% (CodeWiki) versus 3.60\% (DeepWiki) indicate comparable assessment reliability while achieving superior overall quality scores. A pilot human study (Appendix~\ref{app:human-eval}) provides additional validation, showing that human preferences align with automated evaluation results (CodeWiki preferred in 7/9 assessments across 3 participants and 3 repositories).

\textbf{Answer:} CodeWiki's hierarchical decomposition enables consistent scalability across repository sizes from 86K to 1.4M LOC, with performance variations primarily attributable to language-specific characteristics rather than scale limitations. Human evaluation validates alignment between our automated methodology and human judgment.
\section{Conclusion}
We presented CodeWiki, an open-source semi-agentic framework addressing automated repository-level documentation through hierarchical decomposition and recursive processing. CodeWiki performs bottom-up hierarchical synthesis, generating parent modules from children while integrating visual artifacts. Across 7 multilanguage repositories, CodeWiki achieves 68.79\% quality score, outperforming DeepWiki's 64.06\%, demonstrating strong performance across diverse programming languages.

Results show performance variations correlate primarily with language characteristics rather than repository size. High-level languages show consistent improvements across scales, while systems languages face challenges regardless of configuration.

Future directions include: developing specialized parsing modules for improved systems language support, extending to multi-version documentation tracking codebase evolution, and leveraging comprehensive documentation for downstream tasks like code migration. By releasing CodeWiki as open source, we accelerate research and enable the community to build upon our hierarchical processing approach.

\section*{Limitations}
While CodeWiki demonstrates strong performance across multiple languages and repository scales, several limitations warrant discussion.

\paragraph{Systems Programming Languages}
Our framework exhibits reduced effectiveness on systems programming languages (C, C++) compared to high-level languages, achieving 53.24\% versus 79.14\% average scores. This gap stems from the inherent complexity of low-level constructs such as pointer operations, manual memory management, and template metaprogramming. For systems languages these constructs are the architecture, making this a limitation of our current analysis phase rather than a secondary concern. Developing specialized parsing modules that leverage language-specific characteristics represents an important direction for future work.

\paragraph{Rubric Generation}
The automatically generated rubrics, while showing 73.65\% semantic and 70.84\% structural reliability across model families (Appendix~\ref{app:rubric-reliability}), have not undergone comprehensive human verification. Some generated rubrics may be difficult to interpret, and future work should incorporate systematic human validation to ensure rubric quality and clarity.

\paragraph{Evaluation Methodology}
Our evaluation methodology relies on LLM-based assessment with multi-model consensus, which may introduce model-specific biases despite our mitigation strategies. A pilot human study (Appendix~\ref{app:human-eval}) with 3 participants across 3 repositories showed alignment between human preferences and automated scores (CodeWiki preferred in 7/9 assessments), providing preliminary validation. However, broader human evaluation with more participants and systematic rubric-based assessment would further strengthen the benchmark's validity. The average standard deviation of 3.84\% indicates reasonable consistency, yet edge cases may benefit from expert validation.

\ifbeautified
  \clearpage
  \bibliographystyle{plainnat}
\else
  \bibliographystyle{acl_natbib}
\fi
\bibliography{main}

\begin{thebibliography}{63}
\providecommand{\natexlab}[1]{#1}
\providecommand{\url}[1]{\texttt{#1}}
\expandafter\ifx\csname urlstyle\endcsname\relax
  \providecommand{\doi}[1]{doi: #1}\else
  \providecommand{\doi}{doi: \begingroup \urlstyle{rm}\Url}\fi

\bibitem[Ahmed and Devanbu(2023)]{fewshot-code-summarization-ase-2022}
Toufique Ahmed and Premkumar Devanbu.
\newblock Few-shot training llms for project-specific code-summarization.
\newblock In \emph{Proceedings of the 37th IEEE/ACM International Conference on
  Automated Software Engineering}, ASE '22, New York, NY, USA, 2023.
  Association for Computing Machinery.
\newblock ISBN 9781450394758.
\newblock \doi{10.1145/3551349.3559555}.
\newblock URL \url{https://doi.org/10.1145/3551349.3559555}.

\bibitem[Arora et~al.(2025)Arora, Wei, Hicks, Bowman, Quiñonero-Candela,
  Tsimpourlas, Sharman, Shah, Vallone, Beutel, Heidecke, and
  Singhal]{arora2025healthbenchevaluatinglargelanguage}
Rahul~K. Arora, Jason Wei, Rebecca~Soskin Hicks, Preston Bowman, Joaquin
  Quiñonero-Candela, Foivos Tsimpourlas, Michael Sharman, Meghan Shah, Andrea
  Vallone, Alex Beutel, Johannes Heidecke, and Karan Singhal.
\newblock Healthbench: Evaluating large language models towards improved human
  health, 2025.
\newblock URL \url{https://arxiv.org/abs/2505.08775}.

\bibitem[Austin et~al.(2021)Austin, Odena, Nye, Bosma, Michalewski, Dohan,
  Jiang, Cai, Terry, Le, and Sutton]{mbpp}
Jacob Austin, Augustus Odena, Maxwell Nye, Maarten Bosma, Henryk Michalewski,
  David Dohan, Ellen Jiang, Carrie Cai, Michael Terry, Quoc Le, and Charles
  Sutton.
\newblock Program synthesis with large language models, 2021.

\bibitem[Bui et~al.(2023)Bui, Le, Wang, Li, Gotmare, and Hoi]{bui2023codetf}
Nghi~DQ Bui, Hung Le, Yue Wang, Junnan Li, Akhilesh~Deepak Gotmare, and
  Steven~CH Hoi.
\newblock Codetf: One-stop transformer library for state-of-the-art code llm.
\newblock \emph{arXiv preprint arXiv:2306.00029}, 2023.

\bibitem[Cassano et~al.(2023)Cassano, Gouwar, Nguyen, Nguyen, Phipps-Costin,
  Pinckney, Yee, Zi, Anderson, Feldman, Guha, Greenberg, and Jangda]{mul-e}
Federico Cassano, John Gouwar, Daniel Nguyen, Sydney Nguyen, Luna
  Phipps-Costin, Donald Pinckney, Ming-Ho Yee, Yangtian Zi, Carolyn~Jane
  Anderson, Molly~Q Feldman, Arjun Guha, Michael Greenberg, and Abhinav Jangda.
\newblock Multipl-e: A scalable and polyglot approach to benchmarking neural
  code generation.
\newblock \emph{IEEE Transactions on Software Engineering}, 49\penalty0
  (7):\penalty0 3675--3691, 2023.
\newblock \doi{10.1109/TSE.2023.3267446}.

\bibitem[Chen and Huang(2009)]{chenandhuang2009}
Jie-Cherng Chen and Sun-Jen Huang.
\newblock An empirical analysis of the impact of software development problem
  factors on software maintainability.
\newblock \emph{J. Syst. Softw.}, 82\penalty0 (6):\penalty0 981–992, June
  2009.
\newblock ISSN 0164-1212.
\newblock \doi{10.1016/j.jss.2008.12.036}.
\newblock URL \url{https://doi.org/10.1016/j.jss.2008.12.036}.

\bibitem[Chen et~al.(2021)Chen, Tworek, Jun, Yuan, de~Oliveira~Pinto, Kaplan,
  Edwards, Burda, Joseph, Brockman, Ray, Puri, Krueger, Petrov, Khlaaf, Sastry,
  Mishkin, Chan, Gray, Ryder, Pavlov, Power, Kaiser, Bavarian, Winter, Tillet,
  Such, Cummings, Plappert, Chantzis, Barnes, Herbert-Voss, Guss, Nichol,
  Paino, Tezak, Tang, Babuschkin, Balaji, Jain, Saunders, Hesse, Carr, Leike,
  Achiam, Misra, Morikawa, Radford, Knight, Brundage, Murati, Mayer, Welinder,
  McGrew, Amodei, McCandlish, Sutskever, and Zaremba]{chen2021evaluating}
Mark Chen, Jerry Tworek, Heewoo Jun, Qiming Yuan, Henrique~Ponde
  de~Oliveira~Pinto, Jared Kaplan, Harri Edwards, Yuri Burda, Nicholas Joseph,
  Greg Brockman, Alex Ray, Raul Puri, Gretchen Krueger, Michael Petrov, Heidy
  Khlaaf, Girish Sastry, Pamela Mishkin, Brooke Chan, Scott Gray, Nick Ryder,
  Mikhail Pavlov, Alethea Power, Lukasz Kaiser, Mohammad Bavarian, Clemens
  Winter, Philippe Tillet, Felipe~Petroski Such, Dave Cummings, Matthias
  Plappert, Fotios Chantzis, Elizabeth Barnes, Ariel Herbert-Voss,
  William~Hebgen Guss, Alex Nichol, Alex Paino, Nikolas Tezak, Jie Tang, Igor
  Babuschkin, Suchir Balaji, Shantanu Jain, William Saunders, Christopher
  Hesse, Andrew~N. Carr, Jan Leike, Josh Achiam, Vedant Misra, Evan Morikawa,
  Alec Radford, Matthew Knight, Miles Brundage, Mira Murati, Katie Mayer, Peter
  Welinder, Bob McGrew, Dario Amodei, Sam McCandlish, Ilya Sutskever, and
  Wojciech Zaremba.
\newblock Evaluating large language models trained on code, 2021.

\bibitem[Choi et~al.(2023)Choi, Na, Kim, and Lee]{readsum-2023}
Yunseok Choi, Cheolwon Na, Hyojun Kim, and Jee-Hyong Lee.
\newblock Readsum: Retrieval-augmented adaptive transformer for source code
  summarization.
\newblock \emph{IEEE Access}, 11:\penalty0 51155--51165, 2023.
\newblock \doi{10.1109/ACCESS.2023.3271992}.

\bibitem[Crupi et~al.(2025)Crupi, Tufano, Velasco, Mastropaolo, Poshyvanyk, and
  Bavota]{crupi2025effectivenessllmasajudgecodegeneration}
Giuseppe Crupi, Rosalia Tufano, Alejandro Velasco, Antonio Mastropaolo, Denys
  Poshyvanyk, and Gabriele Bavota.
\newblock On the effectiveness of llm-as-a-judge for code generation and
  summarization, 2025.
\newblock URL \url{https://arxiv.org/abs/2507.16587}.

\bibitem[de~Souza et~al.(2005)de~Souza, Anquetil, and
  de~Oliveira]{souza2005astudyofthedocumentation}
Sergio Cozzetti~B. de~Souza, Nicolas Anquetil, and K\'{a}thia~M. de~Oliveira.
\newblock A study of the documentation essential to software maintenance.
\newblock In \emph{Proceedings of the 23rd Annual International Conference on
  Design of Communication: Documenting \& Designing for Pervasive Information},
  SIGDOC '05, page 68–75, New York, NY, USA, 2005. Association for Computing
  Machinery.
\newblock ISBN 1595931759.
\newblock \doi{10.1145/1085313.1085331}.
\newblock URL \url{https://doi.org/10.1145/1085313.1085331}.

\bibitem[{DeepWiki}(2025)]{deepwiki}
{DeepWiki}.
\newblock Deepwiki, 2025.
\newblock URL \url{https://deepwiki.com/}.

\bibitem[Evtikhiev et~al.(2023)Evtikhiev, Bogomolov, Sokolov, and
  Bryksin]{evtikhiev2023outofbleu}
Mikhail Evtikhiev, Egor Bogomolov, Yaroslav Sokolov, and Timofey Bryksin.
\newblock Out of the bleu: How should we assess quality of the code generation
  models?
\newblock \emph{J. Syst. Softw.}, 203\penalty0 (C), September 2023.
\newblock ISSN 0164-1212.
\newblock \doi{10.1016/j.jss.2023.111741}.
\newblock URL \url{https://doi.org/10.1016/j.jss.2023.111741}.

\bibitem[Feng et~al.(2020)Feng, Guo, Tang, Duan, Feng, Gong, Shou, Qin, Liu,
  Jiang, and Zhou]{feng-etal-2020-codebert}
Zhangyin Feng, Daya Guo, Duyu Tang, Nan Duan, Xiaocheng Feng, Ming Gong, Linjun
  Shou, Bing Qin, Ting Liu, Daxin Jiang, and Ming Zhou.
\newblock {C}ode{BERT}: A pre-trained model for programming and natural
  languages.
\newblock In Trevor Cohn, Yulan He, and Yang Liu, editors, \emph{Findings of
  the Association for Computational Linguistics: EMNLP 2020}, pages 1536--1547,
  Online, November 2020. Association for Computational Linguistics.
\newblock \doi{10.18653/v1/2020.findings-emnlp.139}.
\newblock URL \url{https://aclanthology.org/2020.findings-emnlp.139/}.

\bibitem[Guo et~al.(2021)Guo, Ren, Lu, Feng, Tang, Duan, Zhou, and
  Jiang]{guo2021graphcodebert}
Daya Guo, Shuo Ren, Shuai Lu, Zhangyin Feng, Duyu Tang, Nan Duan, Ming Zhou,
  and Daxin Jiang.
\newblock Graphcodebert: Pre-training code representations with data flow.
\newblock In \emph{ICLR}, 2021.

\bibitem[Haldar and Hockenmaier(2024)]{haldar-hockenmaier-2024-analyzing}
Rajarshi Haldar and Julia Hockenmaier.
\newblock Analyzing the performance of large language models on code
  summarization.
\newblock In Nicoletta Calzolari, Min-Yen Kan, Veronique Hoste, Alessandro
  Lenci, Sakriani Sakti, and Nianwen Xue, editors, \emph{Proceedings of the
  2024 Joint International Conference on Computational Linguistics, Language
  Resources and Evaluation (LREC-COLING 2024)}, pages 995--1008, Torino,
  Italia, May 2024. ELRA and ICCL.
\newblock URL \url{https://aclanthology.org/2024.lrec-main.89/}.

\bibitem[Hamel et~al.(2019)Hamel, Wu, Gazit, Allamanis, and
  Brockschmidt]{husain2019codesearchnet}
Husain Hamel, Ho-Hsiang Wu, Tiferet Gazit, Miltiadis Allamanis, and Marc
  Brockschmidt.
\newblock Codesearchnet challenge: Evaluating the state of semantic code
  search.
\newblock In \emph{Proceedings of the 2019 Symposium on Foundations of Software
  Engineering (FSE)}, pages 974--985. ACM, 2019.

\bibitem[He et~al.(2025{\natexlab{a}})He, Shi, Zhuo, Treude, Sun, Xing, Du, and
  Lo]{he2025codecourtroomllmsnew}
Junda He, Jieke Shi, Terry~Yue Zhuo, Christoph Treude, Jiamou Sun, Zhenchang
  Xing, Xiaoning Du, and David Lo.
\newblock From code to courtroom: Llms as the new software judges,
  2025{\natexlab{a}}.
\newblock URL \url{https://arxiv.org/abs/2503.02246}.

\bibitem[He et~al.(2025{\natexlab{b}})He, Treude, and
  Lo]{he2025llmbasedmultiagent}
Junda He, Christoph Treude, and David Lo.
\newblock Llm-based multi-agent systems for software engineering: Literature
  review, vision, and the road ahead.
\newblock \emph{ACM Trans. Softw. Eng. Methodol.}, 34\penalty0 (5), May
  2025{\natexlab{b}}.
\newblock ISSN 1049-331X.
\newblock \doi{10.1145/3712003}.
\newblock URL \url{https://doi.org/10.1145/3712003}.

\bibitem[Hong et~al.(2023)Hong, Zhang, Tang, Jiang, Liu, and
  Yang]{hong2023metagpt}
Xiaohan Hong, Jiaxi Zhang, Wenzhong Tang, Yizhou Jiang, Quan Liu, and Yidong
  Yang.
\newblock Metagpt: Meta programming for multi-agent collaborative framework.
\newblock \emph{arXiv preprint arXiv:2308.00352}, 2023.

\bibitem[Hou et~al.(2023)Hou, Zhao, Liu, Yang, Wang, Li, Luo, Lo, Grundy, and
  Wang]{hou2023large}
Xinyi Hou, Yanjie Zhao, Yue Liu, Zhou Yang, Kailong Wang, Li~Li, Xiapu Luo,
  David Lo, John Grundy, and Haoyu Wang.
\newblock Large language models for software engineering: A systematic
  literature review, 2023.

\bibitem[Khan and Uddin(2023)]{khan2023automaticcodedocumentation}
Junaed~Younus Khan and Gias Uddin.
\newblock Automatic code documentation generation using gpt-3.
\newblock In \emph{Proceedings of the 37th IEEE/ACM International Conference on
  Automated Software Engineering}, ASE '22, New York, NY, USA, 2023.
  Association for Computing Machinery.
\newblock ISBN 9781450394758.
\newblock \doi{10.1145/3551349.3559548}.
\newblock URL \url{https://doi.org/10.1145/3551349.3559548}.

\bibitem[Koreeda et~al.(2023)Koreeda, Morishita, Imaichi, and
  Sogawa]{larch2024}
Yuta Koreeda, Terufumi Morishita, Osamu Imaichi, and Yasuhiro Sogawa.
\newblock Larch: Large language model-based automatic readme creation with
  heuristics.
\newblock In \emph{Proceedings of the 32nd ACM International Conference on
  Information and Knowledge Management}, CIKM '23, page 5066–5070, New York,
  NY, USA, 2023. Association for Computing Machinery.
\newblock ISBN 9798400701245.
\newblock \doi{10.1145/3583780.3614744}.
\newblock URL \url{https://doi.org/10.1145/3583780.3614744}.

\bibitem[LeClair et~al.(2019)LeClair, Jiang, and McMillan]{leclair2019neural}
Alexander LeClair, Siyuan Jiang, and Collin McMillan.
\newblock A neural model for generating natural language summaries of program
  subroutines.
\newblock In \emph{2019 IEEE/ACM 41st International Conference on Software
  Engineering (ICSE)}, pages 795--806. IEEE, 2019.

\bibitem[Li et~al.(2023)Li, Zhang, Luo, Cai, Fang, and Yuan]{Liyao}
Yao Li, Tao Zhang, Xiapu Luo, Haipeng Cai, Sen Fang, and Dawei Yuan.
\newblock Do pretrained language models indeed understand software engineering
  tasks?
\newblock \emph{IEEE Transactions on Software Engineering}, 49\penalty0
  (10):\penalty0 4639--4655, 2023.
\newblock \doi{10.1109/TSE.2023.3308952}.

\bibitem[Lin et~al.(2025)Lin, Deng, Chandu, Ravichander, Pyatkin, Dziri, Bras,
  and Choi]{lin2025wildbench}
Bill~Yuchen Lin, Yuntian Deng, Khyathi Chandu, Abhilasha Ravichander, Valentina
  Pyatkin, Nouha Dziri, Ronan~Le Bras, and Yejin Choi.
\newblock Wildbench: Benchmarking {LLM}s with challenging tasks from real users
  in the wild.
\newblock In \emph{The Thirteenth International Conference on Learning
  Representations}, 2025.
\newblock URL \url{https://openreview.net/forum?id=MKEHCx25xp}.

\bibitem[Lin(2004)]{lin2004rouge}
Chin-Yew Lin.
\newblock Rouge: A package for automatic evaluation of summaries.
\newblock In \emph{Text summarization branches out}, pages 74--81, 2004.

\bibitem[Liu et~al.(2023{\natexlab{a}})Liu, Zeng, Wang, and
  Liang]{liu2023learninggraphbased}
Jiahao Liu, Jun Zeng, Xiang Wang, and Zhenkai Liang.
\newblock Learning graph-based code representations for source-level functional
  similarity detection.
\newblock In \emph{2023 IEEE/ACM 45th International Conference on Software
  Engineering (ICSE)}, pages 345--357, 2023{\natexlab{a}}.
\newblock \doi{10.1109/ICSE48619.2023.00040}.

\bibitem[Liu et~al.(2023{\natexlab{b}})Liu, Fu, Xie, Chen, Pang, Qian, Ma, and
  Radev]{liu2023geval}
Yang Liu, Yao Fu, Yujie Xie, Xinyi Chen, Bo~Pang, Chenyan Qian, Teng Ma, and
  Dragomir Radev.
\newblock G-eval: Nlg evaluation using gpt-4 with better human alignment.
\newblock \emph{arXiv preprint arXiv:2311.08788}, 2023{\natexlab{b}}.

\bibitem[Lomshakov et~al.(2024)Lomshakov, Podivilov, Savin, Baryshnikov,
  Lisevych, and Nikolenko]{lomshakov-etal-2024-proconsul}
Vadim Lomshakov, Andrey Podivilov, Sergey Savin, Oleg Baryshnikov, Alena
  Lisevych, and Sergey Nikolenko.
\newblock {P}ro{C}on{S}u{L}: Project context for code summarization with
  {LLM}s.
\newblock In Franck Dernoncourt, Daniel Preo{\c{t}}iuc-Pietro, and Anastasia
  Shimorina, editors, \emph{Proceedings of the 2024 Conference on Empirical
  Methods in Natural Language Processing: Industry Track}, pages 866--880,
  Miami, Florida, US, November 2024. Association for Computational Linguistics.
\newblock \doi{10.18653/v1/2024.emnlp-industry.65}.
\newblock URL \url{https://aclanthology.org/2024.emnlp-industry.65/}.

\bibitem[Luo et~al.(2024)Luo, Ye, Liang, Zhang, Qin, Lu, Wu, Cong, Lin, Zhang,
  Che, Liu, and Sun]{luo-etal-2024-repoagent}
Qinyu Luo, Yining Ye, Shihao Liang, Zhong Zhang, Yujia Qin, Yaxi Lu, Yesai Wu,
  Xin Cong, Yankai Lin, Yingli Zhang, Xiaoyin Che, Zhiyuan Liu, and Maosong
  Sun.
\newblock {R}epo{A}gent: An {LLM}-powered open-source framework for
  repository-level code documentation generation.
\newblock In Delia~Irazu Hernandez~Farias, Tom Hope, and Manling Li, editors,
  \emph{Proceedings of the 2024 Conference on Empirical Methods in Natural
  Language Processing: System Demonstrations}, pages 436--464, Miami, Florida,
  USA, November 2024. Association for Computational Linguistics.
\newblock \doi{10.18653/v1/2024.emnlp-demo.46}.
\newblock URL \url{https://aclanthology.org/2024.emnlp-demo.46/}.

\bibitem[Makharev and Ivanov(2025)]{makharev2025beyondfunctionlevel}
Vladimir Makharev and Vladimir Ivanov.
\newblock Code summarization beyond function level.
\newblock pages 153--160, 05 2025.
\newblock \doi{10.1109/LLM4Code66737.2025.00024}.

\bibitem[Mastropaolo et~al.(2024)Mastropaolo, Gousios, Bavota, Oliveto, and
  Russo]{mastropaolo2024evaluating}
Ernesto Mastropaolo, Georgios Gousios, Gabriele Bavota, Rocco Oliveto, and
  Barbara Russo.
\newblock Evaluating code summarization techniques: A new metric and an
  empirical characterization.
\newblock In \emph{Proceedings of the 46th International Conference on Software
  Engineering (ICSE)}, 2024.

\bibitem[McBurney et~al.(2018)McBurney, Jiang, Kessentini, Kraft, Armaly,
  Mkaouer, and McMillan]{mcburneyetal-documentationeffort-2018}
Paul~W. McBurney, Siyuan Jiang, Marouane Kessentini, Nicholas~A. Kraft, Ameer
  Armaly, Mohamed~Wiem Mkaouer, and Collin McMillan.
\newblock Towards prioritizing documentation effort.
\newblock \emph{IEEE Trans. Softw. Eng.}, 44\penalty0 (9):\penalty0 897–913,
  September 2018.
\newblock ISSN 0098-5589.
\newblock \doi{10.1109/TSE.2017.2716950}.
\newblock URL \url{https://doi.org/10.1109/TSE.2017.2716950}.

\bibitem[Nassif and Robillard(2025)]{nassif2025non-linear}
Mathieu Nassif and Martin~P. Robillard.
\newblock Non-linear software documentation with interactive code examples.
\newblock \emph{ACM Trans. Softw. Eng. Methodol.}, 34\penalty0 (2), January
  2025.
\newblock ISSN 1049-331X.
\newblock \doi{10.1145/3702976}.
\newblock URL \url{https://doi.org/10.1145/3702976}.

\bibitem[Nguyen et~al.(2025{\natexlab{a}})Nguyen, Phan, Le~Hai, Doan, Nguyen,
  Pham, and Bui]{nguyencodemmlu}
Dung~Manh Nguyen, Thang~Chau Phan, Nam Le~Hai, Tien-Thong Doan, Nam~V Nguyen,
  Quang Pham, and Nghi~DQ Bui.
\newblock Codemmlu: A multi-task benchmark for assessing code understanding \&
  reasoning capabilities of codellms.
\newblock In \emph{The Thirteenth International Conference on Learning
  Representations}, 2025{\natexlab{a}}.

\bibitem[Nguyen et~al.(2025{\natexlab{b}})Nguyen, Chau, Nguyen, and
  Bui]{nguyen2025agilecoder}
Minh~Huynh Nguyen, Thang~Phan Chau, Phong~X Nguyen, and Nghi~DQ Bui.
\newblock Agilecoder: Dynamic collaborative agents for software development
  based on agile methodology.
\newblock In \emph{2025 IEEE/ACM Second International Conference on AI
  Foundation Models and Software Engineering (Forge)}, pages 156--167. IEEE,
  2025{\natexlab{b}}.

\bibitem[Niu et~al.(2022)Niu, Li, Luo, and Ng]{niu2022deep}
Changan Niu, Chuanyi Li, Bin Luo, and Vincent Ng.
\newblock Deep learning meets software engineering: A survey on pre-trained
  models of source code, 2022.

\bibitem[Papineni et~al.(2002)Papineni, Roukos, Ward, and
  Zhu]{papineni2002bleu}
Kishore Papineni, Salim Roukos, Todd Ward, and Wei-Jing Zhu.
\newblock Bleu: a method for automatic evaluation of machine translation.
\newblock In \emph{Proceedings of the 40th annual meeting of the Association
  for Computational Linguistics}, pages 311--318, 2002.

\bibitem[Phan et~al.(2024)Phan, Nguyen, Nguyen, and Bui]{phan2024hyperagent}
Huy~Nhat Phan, Tien~N Nguyen, Phong~X Nguyen, and Nghi~DQ Bui.
\newblock Hyperagent: Generalist software engineering agents to solve coding
  tasks at scale.
\newblock \emph{arXiv preprint arXiv:2409.16299}, 2024.

\bibitem[Poudel et~al.(2024)Poudel, Cook, Traore, and
  Ameli]{poudel2024documint}
Bibek Poudel, Adam Cook, Sekou Traore, and Shelah Ameli.
\newblock Documint: Docstring generation for python using small language
  models.
\newblock \emph{arXiv preprint arXiv:2405.10243)}, 2024.

\bibitem[Qian et~al.(2023)Qian, Zhang, Pan, Wang, Liu, Zhao, and
  Wen]{qian2023chatdev}
Yuzhang Qian, Zian Zhang, Liang Pan, Peng Wang, Shouyi Liu, Wayne~Xin Zhao, and
  Ji-Rong Wen.
\newblock Chatdev: Revolutionizing software development with ai-collaborative
  agents.
\newblock \emph{arXiv preprint arXiv:2307.07924}, 2023.

\bibitem[Rai et~al.(2022)Rai, Belwal, and
  Gupta]{rai2022areviewonsourcecodedocumentation}
Sawan Rai, Ramesh~Chandra Belwal, and Atul Gupta.
\newblock A review on source code documentation.
\newblock \emph{ACM Trans. Intell. Syst. Technol.}, 13\penalty0 (5), June 2022.
\newblock ISSN 2157-6904.
\newblock \doi{10.1145/3519312}.
\newblock URL \url{https://doi.org/10.1145/3519312}.

\bibitem[Roy et~al.(2021{\natexlab{a}})Roy, Fakhoury, and
  Arnaoudova]{royetal-reassessing-2021}
Devjeet Roy, Sarah Fakhoury, and Venera Arnaoudova.
\newblock Reassessing automatic evaluation metrics for code summarization
  tasks.
\newblock In \emph{Proceedings of the 29th ACM Joint Meeting on European
  Software Engineering Conference and Symposium on the Foundations of Software
  Engineering}, ESEC/FSE 2021, page 1105–1116, New York, NY, USA,
  2021{\natexlab{a}}. Association for Computing Machinery.
\newblock ISBN 9781450385626.
\newblock \doi{10.1145/3468264.3468588}.
\newblock URL \url{https://doi.org/10.1145/3468264.3468588}.

\bibitem[Roy et~al.(2021{\natexlab{b}})Roy, Chakraborty, Ray, and
  Kim]{roy2021reassessing}
Rahul Roy, Saikat Chakraborty, Baishakhi Ray, and Miryung Kim.
\newblock Reassessing automatic evaluation metrics for code summarization
  tasks.
\newblock In \emph{Proceedings of the 29th ACM Joint European Software
  Engineering Conference and Symposium on the Foundations of Software
  Engineering (ESEC/FSE)}, pages 1344--1356, 2021{\natexlab{b}}.

\bibitem[Shin et~al.(2023)Shin, Tang, Mohati, Nayebi, Wang, and
  Hemmati]{shin2023prompt}
Jiho Shin, Clark Tang, Tahmineh Mohati, Maleknaz Nayebi, Song Wang, and Hadi
  Hemmati.
\newblock Prompt engineering or fine tuning: An empirical assessment of large
  language models in automated software engineering tasks, 2023.

\bibitem[Sirdeshmukh et~al.(2025)Sirdeshmukh, Deshpande, Mols, Jin, Cardona,
  Lee, Kritz, Primack, Yue, and
  Xing]{sirdeshmukh2025multichallengerealisticmultiturnconversation}
Ved Sirdeshmukh, Kaustubh Deshpande, Johannes Mols, Lifeng Jin, Ed-Yeremai
  Cardona, Dean Lee, Jeremy Kritz, Willow Primack, Summer Yue, and Chen Xing.
\newblock Multichallenge: A realistic multi-turn conversation evaluation
  benchmark challenging to frontier llms, 2025.
\newblock URL \url{https://arxiv.org/abs/2501.17399}.

\bibitem[{Stack Overflow}(2025)]{stackoverflow2025}
{Stack Overflow}.
\newblock 2025 stack overflow developer survey: Ai, 2025.
\newblock URL \url{https://survey.stackoverflow.co/2025/ai}.
\newblock Based on 49,000+ developer responses.

\bibitem[Starace et~al.(2025)Starace, Jaffe, Sherburn, Aung, Chan, Maksin,
  Dias, Mays, Kinsella, Thompson, Heidecke, Glaese, and
  Patwardhan]{starace2025paperbenchevaluatingaisability}
Giulio Starace, Oliver Jaffe, Dane Sherburn, James Aung, Jun~Shern Chan, Leon
  Maksin, Rachel Dias, Evan Mays, Benjamin Kinsella, Wyatt Thompson, Johannes
  Heidecke, Amelia Glaese, and Tejal Patwardhan.
\newblock Paperbench: Evaluating ai's ability to replicate ai research, 2025.
\newblock URL \url{https://arxiv.org/abs/2504.01848}.

\bibitem[Sun et~al.(2023)Sun, Fang, You, Miao, Liu, Li, Deng, Huang, Chen,
  Zhang, et~al.]{sun2023automatic}
Weisong Sun, Chunrong Fang, Yudu You, Yun Miao, Yi~Liu, Yuekang Li, Gelei Deng,
  Shenghan Huang, Yuchen Chen, Quanjun Zhang, et~al.
\newblock Automatic code summarization via chatgpt: How far are we?
\newblock \emph{arXiv preprint arXiv:2305.12865}, 2023.

\bibitem[Treude et~al.(2020)Treude, Middleton, and
  Atapattu]{treude2020beyondaccuracy}
Christoph Treude, Justin Middleton, and Thushari Atapattu.
\newblock Beyond accuracy: assessing software documentation quality.
\newblock In \emph{Proceedings of the 28th ACM Joint Meeting on European
  Software Engineering Conference and Symposium on the Foundations of Software
  Engineering}, ESEC/FSE 2020, page 1509–1512, New York, NY, USA, 2020.
  Association for Computing Machinery.
\newblock ISBN 9781450370431.
\newblock \doi{10.1145/3368089.3417045}.
\newblock URL \url{https://doi.org/10.1145/3368089.3417045}.

\bibitem[von~der Mosel et~al.(2023)von~der Mosel, Trautsch, and Herbold]{Mosel}
Julian von~der Mosel, Alexander Trautsch, and Steffen Herbold.
\newblock On the validity of pre-trained transformers for natural language
  processing in the software engineering domain.
\newblock \emph{IEEE Transactions on Software Engineering}, 49\penalty0
  (4):\penalty0 1487--1507, 2023.
\newblock \doi{10.1109/TSE.2022.3178469}.

\bibitem[Wang et~al.(2025)Wang, Guo, Gao, Fan, Chong, and
  Xia]{wang2025canllmsreplacehumanevaluator}
Ruiqi Wang, Jiyu Guo, Cuiyun Gao, Guodong Fan, Chun~Yong Chong, and Xin Xia.
\newblock Can llms replace human evaluators? an empirical study of
  llm-as-a-judge in software engineering.
\newblock \emph{Proc. ACM Softw. Eng.}, 2\penalty0 (ISSTA), June 2025.
\newblock \doi{10.1145/3728963}.
\newblock URL \url{https://doi.org/10.1145/3728963}.

\bibitem[Wang et~al.(2021)Wang, Ren, Lu, Tang, Duan, Zhou, and
  Jiang]{wang2021codet5}
Yue Wang, Shuo Ren, Daya Lu, Duyu Tang, Nan Duan, Ming Zhou, and Daxin Jiang.
\newblock Codet5: Identifier-aware unified pre-trained encoder-decoder models
  for code understanding and generation.
\newblock In \emph{EMNLP}, 2021.

\bibitem[Wang et~al.(2023)Wang, Le, Gotmare, Bui, Li, and
  Hoi]{wang2023codet5plus}
Yue Wang, Hung Le, Akhilesh Gotmare, Nghi Bui, Junnan Li, and Steven Hoi.
\newblock Codet5+: Open code large language models for code understanding and
  generation.
\newblock In \emph{Proceedings of the 2023 conference on empirical methods in
  natural language processing}, pages 1069--1088, 2023.

\bibitem[Wu et~al.(2023)Wu, Liu, Zhang, Li, Wang, Xin, Zhang, Xing, Lu, and
  Liang]{wu2023autogen}
Ziniu Wu, Cheng Liu, Jindong Zhang, Xinyun Li, Yewen Wang, Jimmy Xin, Lianmin
  Zhang, Eric Xing, Yuxin Lu, and Percy Liang.
\newblock Autogen: Enabling next-generation multi-agent communication with
  language models.
\newblock \emph{arXiv preprint arXiv:2309.07864}, 2023.

\bibitem[Yan et~al.(2023)Yan, Liu, Wang, Li, Chen, Wang, Lin, Zhao, Zhu, Deng,
  and Sundaram]{yan2023codescope}
Weixiang Yan, Haitian Liu, Yunkun Wang, Yunzhe Li, Qian Chen, Wen Wang, Tingyu
  Lin, Weishan Zhao, Li~Zhu, Shuiguang Deng, and Hari Sundaram.
\newblock Codescope: An execution-based multilingual multitask multidimensional
  benchmark for evaluating llms on code understanding and generation, 2023.

\bibitem[Yang et~al.(2025)Yang, Simoulin, Qian, Liu, Cao, Teng, and
  Yang]{yang-etal-2025-docagent}
Dayu Yang, Antoine Simoulin, Xin Qian, Xiaoyi Liu, Yuwei Cao, Zhaopu Teng, and
  Grey Yang.
\newblock {D}oc{A}gent: A multi-agent system for automated code documentation
  generation.
\newblock In Pushkar Mishra, Smaranda Muresan, and Tao Yu, editors,
  \emph{Proceedings of the 63rd Annual Meeting of the Association for
  Computational Linguistics (Volume 3: System Demonstrations)}, pages 460--471,
  Vienna, Austria, July 2025. Association for Computational Linguistics.
\newblock ISBN 979-8-89176-253-4.
\newblock \doi{10.18653/v1/2025.acl-demo.44}.
\newblock URL \url{https://aclanthology.org/2025.acl-demo.44/}.

\bibitem[Zhang et~al.(2023{\natexlab{a}})Zhang, Fang, Xie, Zhang, Yang, Sun,
  Yu, and Chen]{Zhang2023ASO}
Quanjun Zhang, Chunrong Fang, Yang Xie, Yaxin Zhang, Yun Yang, Weisong Sun,
  Shengcheng Yu, and Zhenyu Chen.
\newblock A survey on large language models for software engineering.
\newblock \emph{ArXiv}, abs/2312.15223, 2023{\natexlab{a}}.
\newblock URL \url{https://api.semanticscholar.org/CorpusID:266551742}.

\bibitem[Zhang et~al.(2023{\natexlab{b}})Zhang, Wang, Yang, Zhang, and
  Zhang]{zhang2023mapcoder}
Xiaoqing Zhang, Zhirui Wang, Lichao Yang, Wei Zhang, and Yong Zhang.
\newblock Mapcoder: Map-reduce-style code generation with multi-agent
  collaboration.
\newblock \emph{arXiv preprint arXiv:2307.15808}, 2023{\natexlab{b}}.

\bibitem[Zheng et~al.(2023{\natexlab{a}})Zheng, Chiang, Sheng, Zhuang,
  Zhanghao, Zhuang, Lin, Li, Li, Xing, et~al.]{zheng2023judging}
Lianmin Zheng, Wei-Lin Chiang, Ying Sheng, Siyuan Zhuang, Wu~Zhanghao, Yonghao
  Zhuang, Zi~Lin, Zhuohan Li, Dacheng Li, Eric Xing, et~al.
\newblock Judging llm-as-a-judge with mt-bench and chatbot arena,
  2023{\natexlab{a}}.

\bibitem[Zheng et~al.(2023{\natexlab{b}})Zheng, Ning, Chen, Wang, Chen, Guo,
  and Wang]{zheng2023understanding}
Zibin Zheng, Kaiwen Ning, Jiachi Chen, Yanlin Wang, Wenqing Chen, Lianghong
  Guo, and Weicheng Wang.
\newblock Towards an understanding of large language models in software
  engineering tasks, 2023{\natexlab{b}}.

\bibitem[Zhi et~al.(2015)Zhi, Garousi-Yusifo\u{g}lu, Sun, Garousi, Shahnewaz,
  and Ruhe]{zhi2015costbenefitquality}
Junji Zhi, Vahid Garousi-Yusifo\u{g}lu, Bo~Sun, Golara Garousi, Shawn
  Shahnewaz, and Guenther Ruhe.
\newblock Cost, benefits and quality of software development documentation.
\newblock \emph{J. Syst. Softw.}, 99\penalty0 (C):\penalty0 175–198, January
  2015.
\newblock ISSN 0164-1212.
\newblock \doi{10.1016/j.jss.2014.09.042}.
\newblock URL \url{https://doi.org/10.1016/j.jss.2014.09.042}.

\bibitem[Zhuge et~al.(2024)Zhuge, Zhao, Ashley, Wang, Khizbullin, Xiong, Liu,
  Chang, Krishnamoorthi, Tian, Shi, Chandra, and Schmidhuber]{zhuge2024agent}
Mingchen Zhuge, Changsheng Zhao, Dylan Ashley, Wenyi Wang, Dmitrii Khizbullin,
  Yunyang Xiong, Zechun Liu, Ernie Chang, Raghuraman Krishnamoorthi, Yuandong
  Tian, Yangyang Shi, Vikas Chandra, and J{\"u}rgen Schmidhuber.
\newblock Agent-as-a-judge: Evaluate agents with agents.
\newblock \emph{arXiv preprint arXiv:2410.10934}, 2024.

\end{thebibliography}

\ifbeautified
  \clearpage
  \beginappendix
\else
  \newpage
  \newpage
  \appendix
  \label{sec:Appendix}
\fi
\begin{center}
{\bf{\LARGE{Appendices}}}
\end{center}

\appendix

\section{Example Generated Documentation}
\label{app:documentation-example}

\begin{figure*}[!ht]
    \centering
    \includegraphics[width=\textwidth]{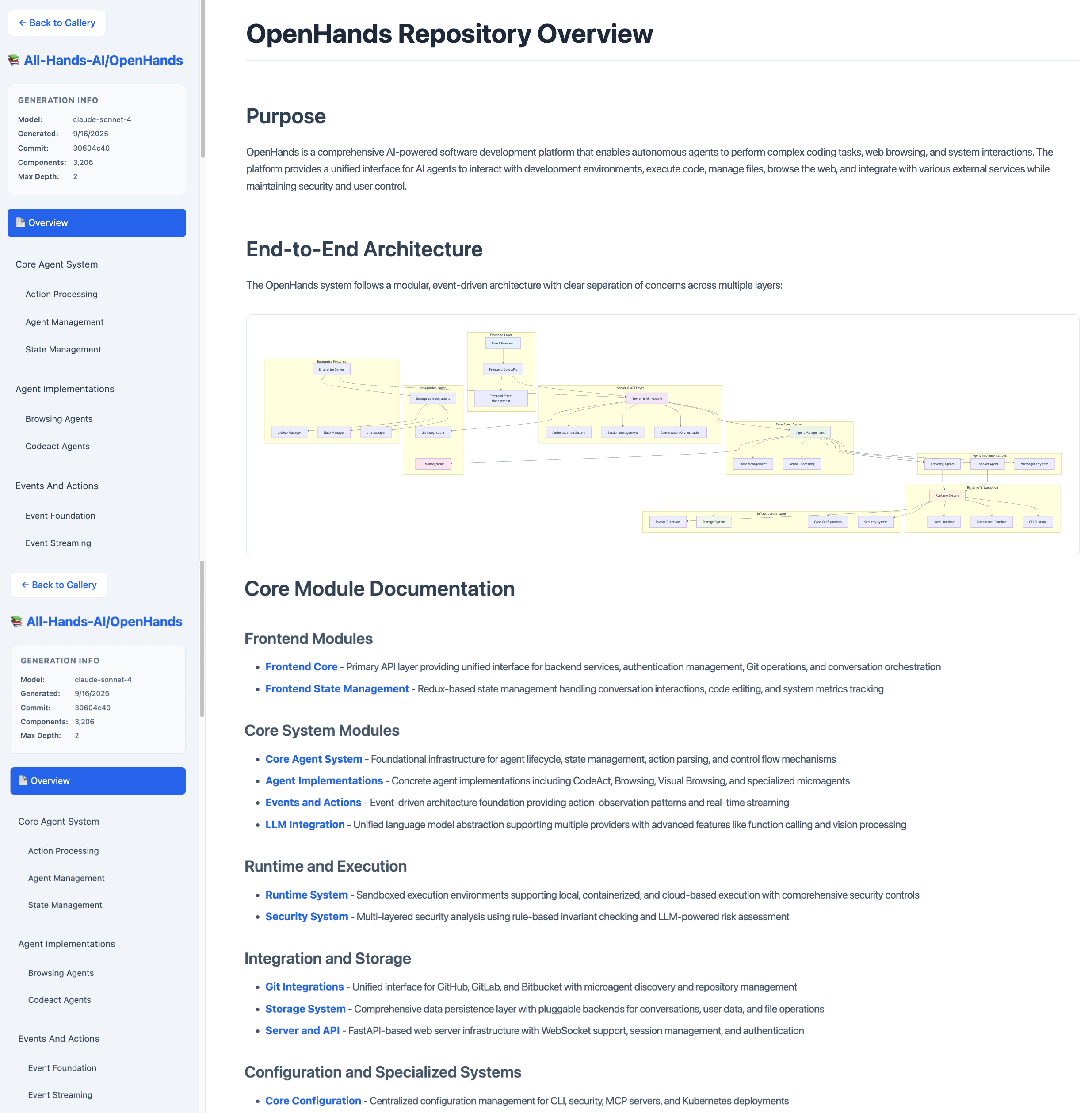}
    \caption{Example repository-level documentation generated by CodeWiki for the All-Hands-AI--OpenHands repository. The documentation includes a comprehensive overview, architectural diagrams showing the modular event-driven architecture, and hierarchical navigation of core components including the agent system, implementations, and event processing modules.}
    \label{fig:openhands-doc-example}
\end{figure*}

Figure~\ref{fig:openhands-doc-example} shows an example of repository-level documentation generated by CodeWiki for the All-Hands-AI--OpenHands repository. The documentation provides a comprehensive overview including the repository's purpose, end-to-end architecture with visual diagrams, and hierarchical navigation of core components. This demonstrates CodeWiki's ability to synthesize both textual explanations and visual artifacts that capture architectural patterns and module relationships, enabling developers to quickly understand the system's structure and design decisions.

\section{Background and Related Work}
\label{app:related-work}

\subsection{Code Summarization and Documentation}

\subsubsection{Existing Approaches}

Advances in natural language processing (NLP) have enabled significant progress in the task of code summarization. Existing methods ~\citep{khan2023automaticcodedocumentation, poudel2024documint, makharev2025beyondfunctionlevel, lomshakov-etal-2024-proconsul,fewshot-code-summarization-ase-2022, readsum-2023,luo-etal-2024-repoagent, yang-etal-2025-docagent} primarily focus on function-level documentation, and can broadly be divided into two categories based on the type of context they exploit (see Table~\ref{tab:code_summ_methods} for a comprehensive comparison):

\paragraph{Local-Level Context}
Early neural approaches established the foundation for automated documentation. CodeBERT~\citep{feng-etal-2020-codebert} demonstrated effective bimodal training on code and natural language, while more recent LLM applications have shown substantial improvements. For instance,~\citep{khan2023automaticcodedocumentation} reported 20.6 BLEU scores across six languages using GPT-3, an 11.2\% improvement over prior methods. DocuMint~\citep{poudel2024documint} further explored the use of small language models for Python docstring generation with promising results. However, these approaches primarily rely on local code context and do not leverage repository-level information to improve documentation quality.

\paragraph{Codebase-Level Context}
Recognizing the limitations of function-only context, recent work has explored broader codebase information to enrich documentation. ~\citep{makharev2025beyondfunctionlevel} demonstrated that incorporating class- and repository-level context can substantially improve summary quality through few-shot learning and retrieval-augmented generation. ProConSuL~\citep{lomshakov-etal-2024-proconsul} incorporated project-level information via compiler analysis and call graphs, employing a two-phase training procedure to reduce hallucinations. RepoAgent~\citep{luo-etal-2024-repoagent} introduced comprehensive LLM-powered repository documentation, though it was limited in content diversity and architectural depth. DocAgent~\citep{yang-etal-2025-docagent} advanced the field through multi-agent collaboration with topological processing, employing specialized agents that respect code dependencies to improve completeness and accuracy.

$\hookrightarrow$ \textit{Limitations and Our Motivation: } Despite this progress, most approaches still target isolated snippets (e.g., functions or classes) and thus fail to capture the underlying architecture of the repository, as highlighted in Table~\ref{tab:code_summ_methods}. For a developer onboarding a completely new codebase, the primary challenge is not understanding what each individual function or class does, but rather grasping the high-level overview of the system, its dataflow and how different components interact. This underscores the need for an automatic methodology capable of generating documentation that spans from low-level details to high-level repository-wide perspectives, covering critical features and providing visual representations that support developer comprehension. Our approach, CodeWiki, addresses these limitations by generating true repository-level documentation while leveraging full codebase context and supporting multiple programming languages.

\subsection{Repository-Level Documentation Systems}

Several systems have addressed repository-level documentation generation with varying approaches and scopes:

\paragraph{Open-Source Alternatives}
\textbf{OpenDeepWiki} and \textbf{deepwiki-open} are community-developed tools that apply LLMs to entire repositories. However, as shown in Table~\ref{tab:results}, these approaches exhibit degraded performance on larger codebases (averaging 47.13\% and 50.05\% respectively), suggesting that simpler whole-repository prompting does not scale effectively to complex projects.

\textbf{RepoAgent}~\citep{luo-etal-2024-repoagent} aggregates function and class-level documentation into repository structures using dependency analysis. While effective for component cataloging, it does not perform hierarchical synthesis where parent modules summarize and contextualize their children.

\textbf{codebase2tutorial}\footnote{\url{https://code2tutorial.com/}} generates tutorial-style documentation using whole-repository prompting, focusing on educational walkthroughs rather than comprehensive architectural documentation.

\textbf{LARCH}~\citep{larch2024} targets README generation specifically, addressing a narrower scope than full repository documentation.

\textbf{Agent-as-a-Judge}~\citep{zhuge2024agent} includes an OpenWiki feature that leverages existing READMEs and applies LLM assessment, representing a different methodology focused on evaluation rather than generation.

\paragraph{CodeWiki's Distinguishing Characteristics}
CodeWiki differs from these approaches through: (1) true hierarchical synthesis where parent modules are generated by summarizing and contextualizing child documentation, (2) integrated visual artifacts (architecture diagrams, data flow visualizations) generated alongside textual content, (3) dynamic delegation enabling scalable processing of arbitrarily large repositories, and (4) cross-module reference management maintaining documentation coherence across module boundaries.

\subsubsection{Existing Evaluation Methods}

Most existing works evaluate the quality of generated documentation using textual similarity metrics such as BLEU or ROUGE. These metrics measure the overlap between generated text and reference ground-truth documentation in terms of lexical or token-level similarity.

$\hookrightarrow$ \textit{Limitations and Our Motivation:} While these methods provide a lightweight quantitative measure, they often fail to capture deeper semantic aspects of the documentation, such as clarity, correctness, and completeness in explaining the underlying code behavior. Recent studies~\citep{royetal-reassessing-2021, evtikhiev2023outofbleu} have shown that traditional metrics like BLEU and ROUGE do not consistently correlate with human judgment of documentation quality. This gap highlights the need for an assessment system that can fairly and accurately evaluate documentation, ensuring that the generated content truly supports developers in comprehending the system.

\begin{table}[!t]
\centering
\caption{Comparison of different code summarization approaches. Repo-Level Doc. indicates whether the approach generates repository-level documentation output. Codebase Context specifies whether the codebase information is used as input to generate the documentation.}
\label{tab:code_summ_methods}
\begin{adjustbox}{width=\columnwidth}
\begin{tabular}{lccc}
\hline
\textbf{Approach} & \textbf{Repo-Level Doc.} & \textbf{Codebase Context} & \textbf{Multilingual} \\
\hline
DocuMint & \xmark & \xmark & \xmark \\
ASAP & \xmark & \cmark & \cmark \\
ProConSul & \xmark & \cmark & \xmark \\
DocAgent & \xmark & \cmark & \xmark \\
RepoAgent & \xmark & \cmark & \xmark \\
\textbf{CodeWiki} & \cmark & \cmark & \cmark \\
\hline
\end{tabular}
\end{adjustbox}
\end{table}

\subsection{Rubric-based evaluation}
A repository can be described through multiple representations, each using different structures and presentation styles while sharing the same underlying architecture. This diversity makes it difficult to evaluate open-ended outputs automatically, as no single metric can reliably evaluate which version is better. Although human evaluation provides trustworthy results, it is resource-intensive and slow. To overcome these limitations and facilitate rapid assessment, many benchmarks \citep{starace2025paperbenchevaluatingaisability,lin2025wildbench, sirdeshmukh2025multichallengerealisticmultiturnconversation, arora2025healthbenchevaluatinglargelanguage} often employ large language models (LLMs) as automated evaluators, using task-specific rubrics to judge the quality of generated outputs. In our CodeWikiBench benchmark, we similarly construct a set of hierarchical rubrics based on the official documentation, providing a structured and fine-grained framework to evaluate the quality and completeness of generated repository-level documentation.

\section{Algorithm Details}
\label{app:algorithm}

\begin{algorithm}[h]
\caption{Recursive Module Documentation}
\label{alg:recursive-doc}
\KwIn{Module tree $T$, dependency graph $G$}
\KwOut{Complete repository documentation}
Initialize documentation workspace\;
\For{each leaf module $m$ in $T$}{
    agent $\leftarrow$ CreateAgent($m$, $G$, $T$)\;
    \While{$m$ not fully processed}{
        agent.GenerateDoc($m$.name)\;
        \If{delegation requested}{
            sub $\leftarrow$ agent.GetDelegationSpecs()\;
            UpdateModuleTree($T$, $m$, sub)\;
            \ForEach{new leaf $m'$ in sub}{
                Recursively process $m'$\;
            }
            ReviseDoc($m$.name)\;
        }
    }
}
ProcessParentModules($T$)\;
\Return Repository documentation\;
\end{algorithm}

\section{Rubric Reliability Assessment}
\label{app:rubric-reliability}

To ensure the quality and trustworthiness of automatically generated evaluation rubrics, we developed a comprehensive reliability assessment framework that quantifies the consistency and structural integrity of rubrics produced by different language models. This assessment methodology provides empirical measures of rubric reliability across multiple dimensions.

Given a set of $n$ rubrics $\{R_1, R_2, \ldots, R_n\}$ produced by different models for the same documentation corpus, we compute pairwise consistency scores across both semantic and structural dimensions.

\subsection{Semantic Consistency Analysis} 

We evaluate semantic consistency by extracting all requirement texts from each rubric and compute semantic similarity using distributed representations. Let $T_i = \{t_1^{(i)}, t_2^{(i)}, \ldots, t_{m_i}^{(i)}\}$ denote the set of requirement texts extracted from rubric $R_i$, where $m_i$ represents the total number of requirements in $R_i$.

For each requirement text $t_k^{(i)} \in T_i$, we identify its best semantic match in $T_j$ using cosine similarity between embeddings:
\begin{equation}
\text{sim}_{\text{best}}(t_k^{(i)}, T_j) = \max_{t_\ell^{(j)} \in T_j} \cos(\mathbf{e}(t_k^{(i)}), \mathbf{e}(t_\ell^{(j)}))
\end{equation}

where $\mathbf{e}(\cdot)$ represents the embedding function and $\cos(\cdot, \cdot)$ denotes cosine similarity. The bidirectional semantic similarity between rubrics is then:
\begin{align}
\text{Sim}_{\text{semantic}}(R_i, R_j) 
&= \frac{1}{m_i + m_j} \Bigg( 
    \sum_{k=1}^{m_i} \text{sim}_{\text{best}}(t_k^{(i)}, T_j) \nonumber \\
&\quad + \sum_{\ell=1}^{m_j} \text{sim}_{\text{best}}(t_\ell^{(j)}, T_i) 
\Bigg)
\end{align}

This formulation ensures that both the coverage of requirements from $R_i$ in $R_j$ and vice versa are considered, providing a balanced measure of semantic overlap.

\subsection{Structural Consistency Analysis} 

Structural consistency evaluates the architectural similarity between rubrics, focusing on hierarchical organization rather than content semantics. We define a comprehensive set of structural features for each rubric:
\begin{itemize}
    \item \textbf{Maximum depth} ($d$): The deepest level in the rubric hierarchy
    \item \textbf{Total items} ($N$): The total number of rubric items across all levels
    \item \textbf{Weight distribution} ($W$): The frequency distribution of importance weights
\end{itemize}

The structural similarity between rubrics $R_i$ and $R_j$ combines multiple normalized metrics:

\begin{equation}
\text{Sim}_{\text{depth}}(R_i, R_j) = 1 - \frac{|d_i - d_j|}{\max(d_i, d_j, 1)}
\end{equation}

\begin{equation}
\text{Sim}_{\text{items}}(R_i, R_j) = 1 - \frac{|N_i - N_j|}{\max(N_i, N_j, 1)}
\end{equation}

For weight distribution similarity, we compute the overlap between normalized probability distributions:
\begin{equation}
\text{Sim}_{\text{weights}}(R_i, R_j) = \sum_{w \in W_i \cup W_j} \min(P_i(w), P_j(w))
\end{equation}

where $P_i(w)$ represents the normalized frequency of weight $w$ in rubric $R_i$.

The overall structural similarity is computed as:
\begin{align}
\text{Sim}_{\text{structural}}(R_i, R_j) 
&= \frac{1}{3}\Big( 
    \text{Sim}_{\text{depth}}(R_i, R_j) \nonumber \\
&\quad + \text{Sim}_{\text{items}}(R_i, R_j) \\
&\quad + \text{Sim}_{\text{weights}}(R_i, R_j) 
\Big)
\end{align}

\subsection{Overall Reliability Score}

The overall reliability score aggregates consistency measures across all model pairs. For $n$ models, we compute $\binom{n}{2}$ pairwise comparisons and derive summary statistics:

\begin{equation}
\text{Score}^{semantic}_{\text{reliability}} = \frac{1}{\binom{n}{2}} \sum_{i=1}^{n-1} \sum_{j=i+1}^{n} \text{Sim}_{\text{semantic}}(R_i, R_j)
\end{equation}
\begin{equation}
\text{Score}^{structural}_{\text{reliability}} = \frac{1}{\binom{n}{2}} \sum_{i=1}^{n-1} \sum_{j=i+1}^{n} \text{Sim}_{\text{structural}}(R_i, R_j)
\end{equation}

Table~\ref{tab:rubric_reliability} presents the reliability assessment results across our benchmark repositories, demonstrating consistent rubric generation with average semantic reliability of 73.65\% and structural reliability of 70.84\%.

\begin{table}[!ht]
\centering
\caption{Rubric Reliability Assessment: Semantic and Structural Consistency Scores}
\label{tab:rubric_reliability}
\begin{adjustbox}{width=\columnwidth}
\begin{tabular}{lcc}
\hline
\textbf{Repository} & \textbf{Semantic (\%)} & \textbf{Structural (\%)} \\
\hline
All-Hands-AI--OpenHands  & 73.64 & 68.82 \\
sveltejs--svelte     & 75.56 & 68.88 \\
puppeteer--puppeteer  & 72.97 & 76.36 \\
Unity-Technologies--ml-agents  & 75.95 & 71.22 \\
elastic--logstash   & 73.13 & 68.22 \\
wazuh--wazuh      & 72.20 & 67.62 \\
electron--electron   & 72.12 & 74.79 \\
\hline
\textbf{Average} & \textbf{73.65} & \textbf{70.84} \\
\hline
\end{tabular}
\end{adjustbox}
\end{table}

\section{Temporal Separation Details}
\label{app:temporal-separation}

Table~\ref{tab:data-leakage} presents the specific commit information for each benchmark repository, demonstrating temporal separation from model training data.

\begin{table}[!ht]
\centering
\caption{Repository commit information demonstrating temporal separation from model training data.}
\label{tab:data-leakage}
\begin{adjustbox}{width=\columnwidth}
\begin{tabular}{llcc}
\hline
\textbf{Repository} & \textbf{Language} & \textbf{Commit ID} & \textbf{Commit Date} \\
\hline
OpenHands & Python & 30604c4 & Sep 13, 2025 \\
svelte & JavaScript & be645b4 & Aug 31, 2025 \\
puppeteer & TypeScript & c1105f1 & Aug 29, 2025 \\
ml-agents & C\# & 4cf2f49 & Aug 15, 2025 \\
logstash & Java & 895cfa5 & Aug 29, 2025 \\
wazuh & C & 44b7cd3 & Aug 30, 2025 \\
electron & C++ & 828fd59 & Aug 30, 2025 \\
\hline
\end{tabular}
\end{adjustbox}
\end{table}

\section{Human Evaluation Pilot Study}
\label{app:human-eval}

To provide additional validation of our automated evaluation methodology beyond multi-model consensus, we conducted a pilot human study comparing CodeWiki and DeepWiki documentation outputs.

\subsection{Study Design}

\paragraph{Participants}
We recruited three participants with diverse professional backgrounds relevant to documentation consumption:
\begin{itemize}
    \item \textbf{Business Analyst}: Represents stakeholders seeking high-level architectural understanding
    \item \textbf{AI Researcher}: Represents technical users evaluating documentation for research purposes
    \item \textbf{Software Engineer}: Represents developers using documentation for implementation guidance
\end{itemize}

None of the participants were prior contributors to the evaluated repositories, simulating the common scenario where developers approach unfamiliar codebases.

\paragraph{Evaluation Protocol}
For each of three repositories (OpenHands, svelte, puppeteer), participants:
\begin{enumerate}
    \item Received access to the official codebase and its official documentation
    \item Reviewed anonymized documentation outputs from both CodeWiki and DeepWiki (presentation order randomized)
    \item Assessed which documentation better reflected: (a) correctness relative to official documentation, and (b) coverage of actual repository architecture
    \item Provided a binary preference judgment
\end{enumerate}

The evaluation was conducted blind, with system identities hidden from participants.

\subsection{Results}

\begin{table}[!ht]
\centering
\caption{Pilot human evaluation results showing participant preferences between CodeWiki (CW) and DeepWiki (DW) across three repositories.}
\label{tab:human-eval-detailed}
\begin{tabular}{lccc|c}
\hline
\textbf{Repository} & \textbf{BA} & \textbf{AI} & \textbf{SE} & \textbf{CW Pref.} \\
\hline
OpenHands & CW & CW & CW & 3/3 \\
svelte & CW & CW & DW & 2/3 \\
puppeteer & CW & DW & CW & 2/3 \\
\hline
\textbf{Total} & 3/3 & 2/3 & 2/3 & \textbf{7/9 (77.8\%)} \\
\hline
\end{tabular}
\end{table}

Table~\ref{tab:human-eval-detailed} presents the evaluation results. CodeWiki was preferred in 7 out of 9 assessments (77.8\%). Key observations:

\paragraph{Alignment with Automated Scores}
Human preferences correlate with the magnitude of automated evaluation improvements:
\begin{itemize}
    \item \textbf{OpenHands} (+9.41\% automated improvement): Unanimous CodeWiki preference (3/3)
    \item \textbf{puppeteer} (+18.54\% automated improvement): Strong CodeWiki preference (2/3)
    \item \textbf{svelte} (+3.45\% automated improvement): Closer preferences (2/3 for CodeWiki)
\end{itemize}

This alignment between human judgment and automated evaluation supports the validity of our LLM-based assessment methodology.

\subsection{Limitations}

We acknowledge several limitations of this pilot study:
\begin{itemize}
    \item \textbf{Scale}: Three participants and three repositories provide preliminary validation but not statistical significance
    \item \textbf{Simplified Protocol}: Due to budget constraints, we used preference judgment rather than full rubric-based evaluation
    \item \textbf{Repository Selection}: The evaluated repositories may not represent the full diversity of our benchmark
\end{itemize}

Comprehensive human evaluation with larger participant pools, more repositories, and detailed rubric-based assessment represents important future work to further validate the benchmark methodology.

\section{Implementation Details}
\label{app:implementation}

\subsection{Model Configurations}

\paragraph{Documentation Generation Agent}
\begin{itemize}
    \item Model: Claude Sonnet 4 (claude-sonnet-4-20250514)
    \item Temperature: 0.0
    \item Max tokens: 16384
    \item Context window: 200K tokens
\end{itemize}

\paragraph{Rubric Generation Agents}
\begin{itemize}
    \item Model 1: Claude Sonnet 4 (Temperature: 0.0)
    \item Model 2: Gemini 2.5 Pro (Temperature: 0.0)
    \item Model 3: Kimi K2 Instruct (Temperature: 0.0)
    \item Max tokens per generation: 32768
\end{itemize}

\paragraph{Judge Agents}
\begin{itemize}
    \item Judge 1: Gemini 2.5 Flash (Temperature: 0.0)
    \item Judge 2: GPT OSS 120B (Temperature: 0.0)
    \item Judge 3: Kimi K2 Instruct (Temperature: 0.0)
    \item Binary scoring: 0 (not satisfied) or 1 (satisfied)
\end{itemize}

\subsection{Processing Parameters}

\begin{itemize}
    \item Module decomposition threshold: 32768 tokens per leaf module
    \item Maximum delegation depth: 3 levels
\end{itemize}

\end{document}